\documentclass{IEEEtran}
\usepackage[utf8]{inputenc}
\usepackage{amsthm,authblk,color,booktabs}
\usepackage{graphicx,epstopdf}
\usepackage{epsfig}
\usepackage{subfigure}
\usepackage{amsfonts}
\usepackage{bbold}
\usepackage{cite}
\usepackage{multirow}
\usepackage{amsmath}
\usepackage{url}
\usepackage{tikz}
\usepackage{pstricks,psfrag,pstool}
\usetikzlibrary{shapes.geometric, arrows}
\usetikzlibrary{plotmarks}

\title{Joint Modeling of Received Power, Mean Delay, and Delay Spread for Wideband Radio Channels}
\author{Ayush Bharti, Ramoni Adeogun, Xuesong Cai, Wei Fan, Fran\c{c}ois-Xavier Briol, Laurent Clavier, Troels Pedersen \thanks{Ayush Bharti, Ramoni Adeogun, Xuesong Cai, Wei Fan, and Troels Pedersen are with the Department of Electronic Systems, Aalborg University, Denmark (e-mail: \{ayb, ra, xuc, wfa, troels\}@es.aau.dk). 

Fran\c{c}ois-Xavier Briol is affiliated with the Department of Statistical Science at University College London and the Data-Centric Engineering Programme at The Alan Turing Institute, London, United Kingdom (email: f.briol@ucl.ac.uk). 

Laurent Clavier is with IMT Lille Douai, University of Lille, CNRS, UMR 8520, and F-59000 Lille, France (e-mail: laurent.clavier@imt-lille-douai.fr).}}

\begin{document}

\maketitle

\begin{abstract}
    We propose a multivariate log-normal distribution to jointly model received power, mean delay, and root mean square (rms) delay spread of wideband radio channels, referred to as the standardized temporal moments. The model is validated using experimental data collected from five different measurement campaigns (four indoor and one outdoor scenario). We observe that the received power, the mean delay, and the rms delay spread are correlated random variables, and therefore, should be simulated jointly. Joint models are able to capture the structure of the underlying process, unlike the independent models considered in the literature. The proposed model of the multivariate log-normal distribution is found to be a good fit for a large number of wideband data-sets.
\end{abstract}

\vskip0.5\baselineskip
\section{Introduction}
Standardized temporal moments such as received power, mean delay, and root mean square (rms) delay spread are widely used to summarize power-delay profiles (PDPs) of wideband radio channels. Characterization of these temporal moments is imperative for understanding the effects of multipath propagation on the received signal \cite{Goldsmith2005}, and hence, for the design and analysis of communication and localization systems. The standardized temporal moments are derived from transformations of the raw temporal moments of the instantaneous power of the received signal. Therefore, the raw moments, and consequently the standardized moments, are dependent random variables. The raw temporal moments have recently been used to estimate parameters of stochastic radio channel models from measurements \cite{Wu2008, AyushURSI, AyushSPAWC19, AyushABC, Bharti2020, RamoniMLAWPL}. Mean delay and rms delay spread have also been used to fit an extension of the WINNER~II model to measurements \cite{Latinovic2020}. In applications where multiple temporal moments are used, it can be valuable to consider their dependencies to avoid biases which can occur due to false assumptions of independence.

\begin{figure}
	 \centering
		 \includegraphics[width = 0.38\textwidth]{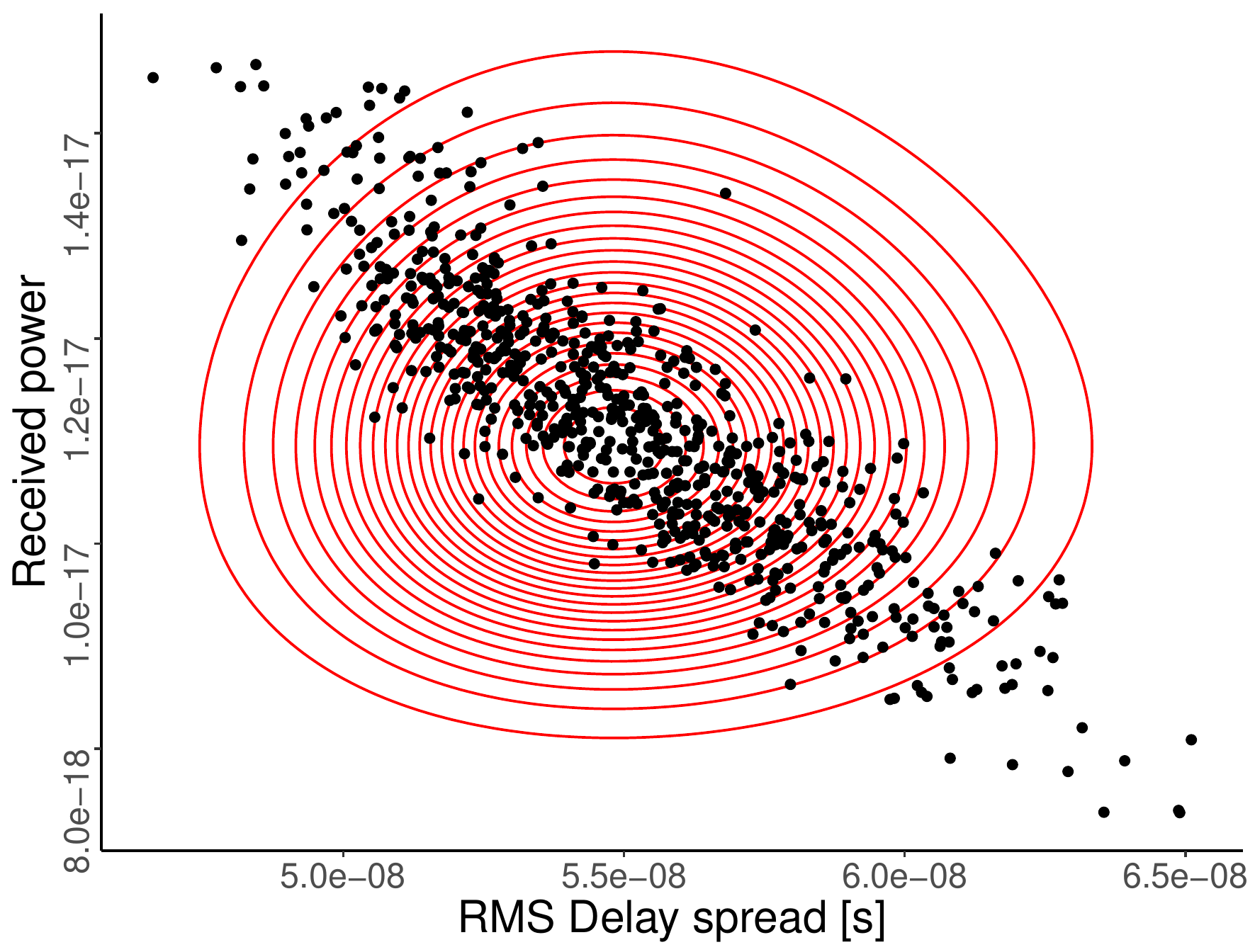}
            \vspace{-2ex}
		 \includegraphics[trim={0 0 0 40}, clip, width = 0.4\textwidth]{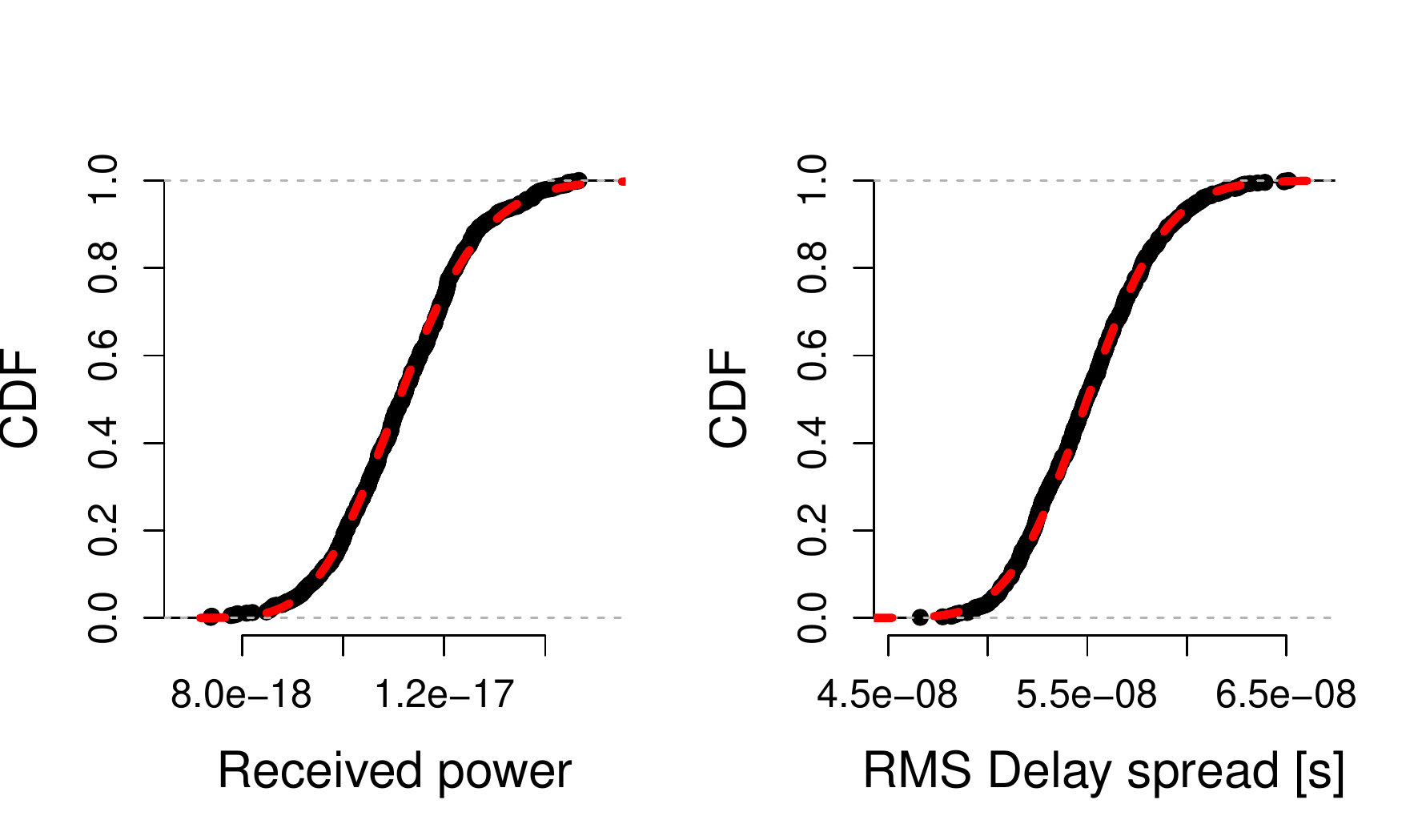}
		   \caption{Scatter plot of received power and rms delay spread obtained from AAU-Hall measurements (see Sec.~\ref{subsec:aauHall}) is shown in black (above). The contour lines from independently fitting log-normal distribution to the measurements is shown in red. The empirical CDFs of the marginals is also shown with the fitted log-normal CDF in red (below). Note that the received power is unitless.}
  \label{fig:intro}
 \end{figure}

Independent modeling of received power, mean delay, and rms delay spread is prevalent in the literature, with their empirical averages and cumulative distributions functions (CDFs) being reported frequently while disregarding their dependencies. A survey of the empirical data available for the delay properties of indoor radio channel is given in \cite{Awad2008}, where a variety of marginal models is fit to the mean delay and rms delay spread from the various data-sets. They obtained log-normal, Gaussian, and Weibull as the best fit models. Empirical distribution of delay spread has been modeled using a log-normal distribution in the 910 MHz channel \cite{Cox1975, Greenstein1997}, the 30 MHz to 400 MHz frequency band \cite{Fischer2013}, at 460~MHz \cite{Wang2019}, at 11~GHz \cite{Li2019}, and at 39~GHz \cite{Zhang2019}. A Gaussian distribution for the rms delay spread has also been proposed based on empirical data in \cite{Yu2017} and \cite{Schmieder2020}. Recently, the rms delay spread has also been modeled using a bimodal Gaussian mixture distribution \cite{Yu2020} and neural networks \cite{Yu2020a}.

The shortcomings of independent modeling become clear by considering jointly the received power and rms delay spread as done in the example in Fig.~\ref{fig:intro}. It is apparent that by fitting independent log-normal models to the received power and the rms delay spread, the marginals of the data are modeled correctly.
However, the correlation information in the data is lost on modeling them independently. Delay spread is previously found to be correlated to received power at 60~GHz \cite{Prokes2019} and to mean delay in the ultra-high frequency band \cite{Varela2001}. One approach to mitigate this problem is to model the standardized moments jointly. An exception to the independent models is the one proposed by Greenstein et al. \cite{Greenstein1997} where they accounted for the correlation between rms delay spread and shadow fading after analysing a wide range of outdoor measurements, mostly in the 900 MHz frequency band. They argued that rms delay spread is log-normally distributed at a given propagation distance, and proposed a joint log-normal model for path gain\footnote{Greenstein et al. \cite{Greenstein1997} defined path gain as the ratio of received power to transmitted power.} and delay spread with a correlation coefficient of --0.75. However, they did not take mean delay into account. Moreover, the correlation coefficient was based on qualitative analysis of scatter plots and on a single measurement setting. The mutual relations between the means of the raw temporal moments have been modeled in \cite{Steinboeck2013, Pedersen2018, Pedersen2019} for the in-room case, while their joint distribution was not studied. To the best of our knowledge, joint characterization of the temporal moments in the millimetre-wave (mm-wave) band has not been done before.

Potentially, the temporal moments could be modeled jointly using a multivariate distribution such that the model could be fitted to new measurements. Joint modeling of multivariate random variables is considerably more involved than modeling of scalar random variables because the model is required to represent the marginals and the dependency structure in the data at the same time. Only a few univariate probability distribution functions (pdfs) exist that have unique multivariate extensions, such as the multivariate Gaussian, log-normal, and Gamma distributions \cite{Kotz2000}. Copulas \cite{Nelsen2007} can also be used to model the dependency structure between the random variables, especially when the marginal distributions lead to a multivariate distribution that is difficult to handle due to the lack of analytical expression or difficulties to estimate the parameters.

After considering several of these methods, we conclude that the multivariate log-normal is a reasonable choice which provides a good balance between goodness-of-fit and ease of interpretation. Moreover, there is substantial support for log-normality of standardized temporal moments in the literature. In this paper, we propose and validate the multivariate log-normal model using a wide variety of measurements taken in different scenarios and frequency ranges, including both indoor and outdoor settings. Measurement campaigns were conducted at Lund University \cite{Gustafson2016}, University of Lille \cite{Fryziel2002}, and Aalborg University (AAU) \cite{AAUIndustryMea}. We also present mm-wave measurements from one indoor and one outdoor campaign in the 28 GHz to 30 GHz band conducted recently at AAU. We compare the proposed model with the multivariate Gaussian and independent marginal models in terms of the Akaike Information Criterion (AIC). Finally, we investigate the model fits to the raw and standardized temporal moments from the measurements. Preliminary results have been published in the conference publication \cite{AyushEUCAP}.

The paper is organized as follows: Section II describes the raw and standardized temporal moments, and Section III presents the model. In Section IV we compare the proposed model with other modeling choices. Section V and VI compare the model fits to the raw and standardized temporal moments of the measurements, respectively. Finally, the conclusions are outlined in Section VII.
\section{Temporal Moments}
Consider a measurement campaign where the channel transfer function between fixed transmit and receive antennas is recorded using a vector network analyzer (VNA). Sampling the transfer function, $H(f)$, at $N_s$ frequency points in the measurement bandwidth $B$ results in a separation of $\Delta f= B/(N_s - 1)$ between the points. We assume that the measurement noise at the $n^\textup{th}$ frequency point, $W_n$, is additive and independent of the transfer function, $H_n$. Then, the measured frequency-domain signal, $Y_n$, reads
\begin{equation}\label{eq:basicModel}
	Y_n = H_n + W_n, \quad n = 0, 1, \dots, (N_s-1).
\end{equation}
Discrete-frequency, continuous-time inverse Fourier transform gives the $1/\Delta f$-periodic measured time-domain signal
\begin{equation}
    y(t) = \frac{1}{N_s} \sum_{n=0}^{N_s-1} Y_n \exp(j2\pi n \Delta f t).
\end{equation}
Note that $y(t)$ is often referred to as the impulse response despite suffering from limited bandwidth and noise. This terminology is somewhat misleading since strictly speaking the impulse response is the inverse Fourier transform of $H(f)$. For large bandwidth and high signal-to-noise ratio (SNR), $y(t)$ can be used as an approximation to the impulse response in the time interval $[0,1/\Delta f]$, provided that the impulse response decays rapidly enough. To avoid this confusion, we refer to $y(t)$ as the measured signal.

The raw temporal moments are summary statistics of the measured signal $y(t)$, where the $k^\textup{th}$ temporal moment is defined as
\begin{equation}\label{eq:temporalMoments}
	m_k = \int_0^{\frac{1}{\Delta f}} t^k \vert y(t)\vert^2 \text{dt}, \quad k = 0,1,\dots, (K-1).
\end{equation}
Here, a total of $K$ raw temporal moments are computed ``instantaneously" per realization of $y(t)$, giving the $K$-dimensional vector $ \mathbf{m}~=~\left[m_0, m_1, \dots, m_{K-1}\right]^\top $. The raw temporal moments are correlated random variables as they are all derived from the received signal power, $|y(t)|^2$. The $ k^\textup{th} $ temporal moment is measured in $ \text{[second]}^k $.

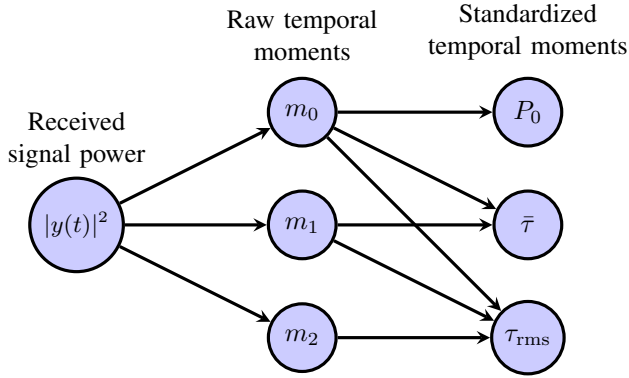
\begin{figure}
  \centering
  \begin{tikzpicture}[roundnode/.style={circle, draw=black, very
      thick, fill=blue!20, minimum size=9mm}, arrow/.style = {very thick,->,>=stealth}]

\node[name=h, roundnode] at (0,0)  {\small $|y(t)|^2$};
\node[name=m0, roundnode] at (3,1.5)  { $m_0$};
\node[name=m1, roundnode] at (3,0)  { $m_1$};
\node[name=m2, roundnode] at (3,-1.5)  { $m_2$};
\node[name=tbar, roundnode] at (6, 0)  { $\bar\tau$};
\node[name=trms, roundnode] at (6,-1.5)  { $\tau_{\mathrm {rms}}$};
\node[name=p0, roundnode] at (6, 1.5)  { $P_0$};
\draw[arrow] (h) --  (m0);
\draw [arrow](h) -- (m1);
\draw [arrow](h) -- (m2);
\draw [arrow](m0) -- (tbar);
\draw [arrow](m1) -- (tbar);
\draw [arrow](m0) -- (trms);
\draw [arrow](m1) -- (trms);
\draw [arrow](m2) -- (trms);
\draw [arrow](m0) -- (p0);
\node [above=0mm,text width=2cm,align=center] at (h.north) {Received signal power};
\node [above=1mm,text width=2.5cm,align=center] at (m0.north)
{Raw temporal moments};
\node [above=1mm,text width=3cm,align=center] at (p0.north)
{Standardized temporal moments};
\end{tikzpicture}
  \caption{The connections between the  magnitude square
    received signal and the summary statistics (raw- and standardized temporal moments).}
  \label{fig:causalRelations}
\end{figure}

The standardized temporal moments are obtained from the first three raw temporal moments. The received power, $P_0$, the mean delay, $\bar \tau$, and the rms delay spread, $ \tau_{\mathrm{rms}} $, are given as
\begin{equation}\label{eq:standardizedMoments}
P_0 = m_0, \quad \bar{\tau} = \frac{m_1}{m_0}, \quad \text{and} \quad \tau_{\mathrm{rms}} = \sqrt{\frac{m_2}{m_0} - \left(\frac{m_1}{m_0}\right)^2}.
\end{equation}
The unit of $ \bar{\tau} $ and $ \tau_{\mathrm{rms}} $ is in seconds whereas $P_0$ is unitless. The deterministic relationship between the raw and the standardized temporal moments is depicted in Fig.~\ref{fig:causalRelations}. The non-linearity of the above transformations and the dependency of the raw temporal moments complicates the joint characterization of mean delay and rms delay spread. Summarizing $ N_{\text{real}} $ realizations of the measured signal into $K$ temporal moments therefore results in the $K \times  N_{\text{real}} $ dimensional matrix, $ \mathbf{M}=\left[\mathbf{m}^{(1)}, \dots, \mathbf{m}^{(N_{\text{real}})}\right] $. We will focus our discussion on the first three temporal moments, $ (m_0,m_1,m_2) $, as they suffice for the received power, mean delay, and rms delay spread but it is straightforward to extend the framework to include more moments as long as the marginal distributions fit the same distribution.

Note that the standardized temporal moments in \eqref{eq:standardizedMoments} are computed from the measured signal, $y(t)$, rather than the channel impulse response. The impulse response is unobservable due to the noise and bandwidth limitations. It is widespread practice to employ a thresholding procedure to reduce the effect of the measurement noise on the estimation of temporal moments. However, such procedures require the setting of a threshold or dynamic range. The choice of the threshold affects the resulting estimates in a manner that makes comparison between measurements obtained with different equipment  difficult. For this reason, we omit any thresholding procedure in the present work.

The finite measurement bandwidth also manifests itself in the rms delay spread as an approximately additive term equal to the delay spread of the transmitted signal. This effect can be partially removed by subtracting the delay spread of the frequency window. This is widespread practice in the literature and results in a good approximation if the bandwidth is large and the SNR is high. However, in case of low SNR and small signal bandwidth, this can lead to inaccurate and sometimes negative estimates of the delay spread. For the measurements considered in Section~\ref{sec:Measurements}, where the bandwidth is very large, the effect of the transmitted signal can be ignored. Hence, we make no attempt to compensate for the effect of a finite measurement bandwidth. 
\section{Proposed Statistical Model}
We intend to jointly model the first three raw temporal moments, $(m_0,m_1,m_2)$, and use the transformation in \eqref{eq:standardizedMoments} to simulate the mean delay and rms delay spread. In principle, the standardized temporal moments could be modeled instead of the raw moments. However, the distribution on the raw moments implies a distribution on the standardized moments from which sampling is straightforward. Modeling the raw moments has the added advantage that their means and covariances are easier to compute analytically for a given channel model than those of the standardized moments due to the non-linear transformation.

We model the vector $ \mathbf{m} = [m_0,m_1,m_2]^\top$ as a multivariate log-normal variable. The exponential of a random vector following a multivariate Gaussian distribution is multivariate log-normal distributed. Let $ \mathbf{x} $ be a $K$-variate normal random vector with mean $ \boldsymbol{\mu} $ and covariance matrix $ \boldsymbol{\Sigma} $. Then its entry-wise exponentiation, $ \mathbf{m} = \exp(\mathbf{x}) $, yields a log-normal vector with pdf
\begin{multline}\label{eq:lognormalPDF}
	f(\mathbf{m};\boldsymbol{\mu}, \boldsymbol{\Sigma})  = \frac{\prod_{k=0}^{K-1} (m_k)^{-1}}{\sqrt{(2\pi)^{K} \det  \boldsymbol{\Sigma}}} \\ \times \exp\left(-\frac{1}{2} (\ln(\mathbf{m}) - \boldsymbol{\mu})^\top \boldsymbol{\Sigma}^{-1} (\ln(\mathbf{m}) - \boldsymbol{\mu} )\right).
\end{multline}
Here the logarithm is taken entry-wise. By property of the marginals of the multivariate Gaussian, it is easy to see that this transform results in a distribution with log-normal marginals. Note that the parameters of a multivariate log-normal are the mean vector and the covariance matrix of the associated multivariate Gaussian distribution. The entries of $ \boldsymbol{\mu} $ and $ \boldsymbol{\Sigma} $ are given as $ \mu_k~=~\mathbb{E}\left[\ln m_k\right]$ and $ \Sigma_{kk'}~=~\text{cov}\left(\ln m_k, \ln m_{k'}\right) $, for $ k,k'~=~0,1,\dots, K-1 $, respectively. Given that raw temporal moments are log-normally distributed, their means and covariances can be related to $\mu$ and $\Sigma$ as
\begin{equation}
	\mathbb{E}\left[m_k\right] = \exp\left(\mu_k + \frac{1}{2}\Sigma_{kk}\right), \quad \text{and}
\end{equation}
\vspace{-4ex}
%
\begin{multline}
    \text{cov}\left(m_k, m_{k'}\right) =  \exp \left(\mu_k + \mu_{k'} + \frac{1}{2}\left(\Sigma_{kk} + \Sigma_{k'k'}\right)\right) \\ \times \left( \exp\left(\Sigma_{kk'}\right) - 1\right).
\end{multline}

Note that we model the raw temporal moments as opposed to Greenstein et al. \cite{Greenstein1997} who model shadow fading and rms delay spread as jointly log-normal. With the proposed model, log-normality is preserved for the received power and mean delay due to the multiplicative transform applied on $m_0$ and $m_1$. However, the distribution of rms delay spread depends on a more complicated transformation (see \eqref{eq:standardizedMoments}) and hence cannot easily be derived in closed form.
\begin{table*}[]
\centering
\caption{Summary of different measurement data-sets.}
\label{table:summary}
\resizebox{\textwidth}{!}{%
\begin{tabular}{@{}cccccccc@{}}
\toprule
       \textbf{Data set}               & \textbf{\begin{tabular}[c]{@{}c@{}}Bandwidth\\ (GHz)\end{tabular}} & \textbf{\begin{tabular}[c]{@{}c@{}}No. of\\ samples\end{tabular}} & \textbf{\begin{tabular}[c]{@{}c@{}}No. of\\ realizations\end{tabular}} & \textbf{\begin{tabular}[c]{@{}c@{}}Antenna\\ Tx/Rx\end{tabular}} & \textbf{\begin{tabular}[c]{@{}c@{}}Dimensions\\ ($\text{m}^3$) \end{tabular}} & \textbf{Scenario} & \textbf{Environment} \\ \midrule
\textbf{Lund Data} \cite{Gustafson2016}    & 58-62                                                              & 801                                                               & 625                                                                    & Biconical/Open waveguide                                              &  $3\times 4 \times 3$                                                                           & NLOS              & Small room           \\
\textbf{Lille Data} \cite{Fryziel2002}   & 59-61                                                              & 1601                                                              & 750                                                                    & Microstrip/Microstrip                                                       &  $5.20 \times 7.15 \times 2.90$                                                                            & LOS               & Large room           \\
\textbf{AAU-Industry} \cite{AAUIndustryMea} & 3-8                                                                & 5001                                                              & 95                                                                     & Biconical/Biconical                                              &  $33 \times 14 \times 6$                                                                            &  Both                 & Industry hall             \\
\textbf{AAU-Hall}     & 28-30                                                              & 1500                                                              &  720                                                                  & Biconical/Biconical                                              & $44\times25\times10$                                                                             &       NLOS             & Large hall           \\
\textbf{AAU-Outdoor}  & 28-30                                                              & 2000                                                              & 360                                                                     & Horn/Biconical                                                   &   ---                                                                            &        LOS          & Outdoor              \\ \bottomrule
\end{tabular}%
}
\end{table*}

\subsection{Estimation of parameters}
The parameters of the proposed model need to be estimated from measured data in order to use the model for simulation purposes. Here, we refer to the matrix of raw temporal moments, $\mathbf M$, as the data. This data matrix is obtained by summarizing $N_{\mathrm{real}}$ realizations of the measured signal using \eqref{eq:temporalMoments}. Assuming independent and identically distributed (iid) realizations, maximum likelihood estimation of $ \boldsymbol{\mu} $ and $ \boldsymbol{\Sigma} $ is achieved by solving the optimization problem,
\begin{equation}\label{eq:ml}
(\hat{\boldsymbol{\mu}}, \hat{\boldsymbol{\Sigma}}) = \underset{\boldsymbol{\mu}, \boldsymbol{\Sigma}}{\mathrm{argmax}} \quad \prod_{i=1}^{N_{\text{real}}} f\left(\mathbf{m}^{(i)}; \boldsymbol{\mu}, \boldsymbol{\Sigma}\right).
\end{equation}
Since $ \boldsymbol{\mu} $ and $ \boldsymbol{\Sigma }$ are the mean vector and the covariance matrix, respectively, of the associated multivariate Gaussian distribution, their maximum likelihood estimates, $ \hat{\boldsymbol{\mu}} $, and $ \hat{\boldsymbol{\Sigma}} $, are
\begin{align}
	\hat{\boldsymbol{\mu}} &= \frac{1}{N_{\text{real}} } \sum_{i=1}^{N_{\text{real} }} \ln \mathbf{m}^{(i)}, \quad \text{and} \label{eq:Muestimates}\\
	\hat{\boldsymbol{\Sigma }} &= \frac{1}{N_{\text{real}} } \sum_{i=1}^{N_{\text{real} }} \left(\ln \mathbf{m}^{(i)} -  \hat{\boldsymbol{\mu}}\right) \left(\ln \mathbf{m}^{(i)} -  \hat{\boldsymbol{\mu}}\right)^\top. \label{eq:Sigmaestimates}
\end{align}
\subsection{Simulation from the model}
Given a particular value of $ \boldsymbol{\mu} $ and $ \boldsymbol{\Sigma} $, simulation from the proposed model is straightforward. To generate one sample of $\mathbf m$, or subsequently, one sample of $(P_0, \bar \tau, \tau_{\mathrm{rms}})$, the following steps should be performed:
\begin{enumerate}
	\item Draw $ \mathbf{x} \sim \mathcal{N}\left(\boldsymbol{\mu}, \boldsymbol{\Sigma}\right) $
	\item Compute entry-wise exponential, $ \mathbf{m} = \exp(\mathbf{x}) $
	\item Compute $ \bar{\tau} $ and $ \tau_{\mathrm{rms}} $ from \eqref{eq:standardizedMoments}
\end{enumerate}
\section{Measurement Data Description}\label{sec:Measurements}
We now describe the different radio channel measurements used to validate the proposed model. An overview of the measurement data-sets is given in Tab.~\ref{table:summary}. The parameter estimates obtained after fitting the proposed model to the measurements are reported in Tab.~\ref{tab:estimates}.
\subsection{Data-set from Lund University}
Polarimetric radio channel measurements at 60 GHz was recorded in a small meeting room of dimensions $3\times 4 \times 3~\text{m}^3$ using a VNA \cite{Gustafson2016}. The room consisted of a table, white-board, bookshelves, and a window on one of the walls. The receive antenna was placed at one corner of the room and the transmit antenna was placed on the table. A water-filled human phantom was used to block the line-of-sight (LOS) path to emulate non-line-of-sight (NLOS) scenario. A $5\times 5$ virtual array of dual-polarized antennas was used with an inter-element spacing of 5 mm at both the transmitter and the receiver. This resulted in a $25\times25$ dual-polarized MIMO system, however, in this paper, we only use the vertical-vertical polarized channels. Measurements were performed in the bandwidth range of 58~GHz to 62~GHz using 801 equally spaced frequency points. For further details on the measurement campaign, see \cite{Gustafson2016}.
\subsection{Data-set from Lille University}
Measurements were taken in a computer laboratory of floor area $7.15\times 5.2 ~\text{m}^2$ at 26 sites, covering the whole room. The 60 GHz channel sounder developed at IEMN \cite{Fryziel2002} used two heterodyne emission and reception heads developed by monolithic integration with frequencies ranging from 57~GHz to 59~GHz and with intermediate frequencies of 1~GHz to 3~GHz. A dedicated network analyzer allows, after calibration, the vectored measure of the frequency transfer function by steps of 1.25~MHz. The resulting impulse response has a delay resolution of 0.5~ns and a maximum measurable delay of 800~ns. In this paper, we select a subset of the entire data-set, specifically, taking the measurements from the first three sites having the same distance between the transmit and receive antennas in LOS condition. Each site consists of 250 positions separated by 2~mm. The transmitter was fixed in a corner, close to the roof, pointed towards the opposite corner. The receiver was oriented towards the transmitter in the horizontal plane but not in the vertical one. Horizontal linear polarization patch antennas were employed.
\subsection{AAU Data, Industry Scenario}
Short-range ultra-wideband measurement campaigns were conducted in a $33 \times 14 \times 6 ~\text{m}^3$ industrial factory hall at the Smart Production Lab, AAU. The factory hall was a typical high clutter density environment with large metallic machinery such as welding machines, hydraulic press, and material processing machines. Measurements were collected over the frequency range 3~GHz to 8~GHz using a Rhode \& Schwarz ZND 8.5~GHz VNA and omni-directional broadband bi-conical antennas at both the transmitter and the receiver. During the measurements, the transmitter was placed at a fixed location and the receiver location was varied to obtain horizontal distances between 1~m and 9~m. A total of 95 channel transfer functions were obtained with a frequency resolution of 1~MHz corresponding to 5001 samples over the 5~GHz bandwidth. Detailed description of the measurements can be found in \cite{AAUIndustryMea}.
\subsection{AAU Data, Hall Scenario} \label{subsec:aauHall}
\begin{figure}
\centering

\subfigure[]{\includegraphics[width=0.5\textwidth]{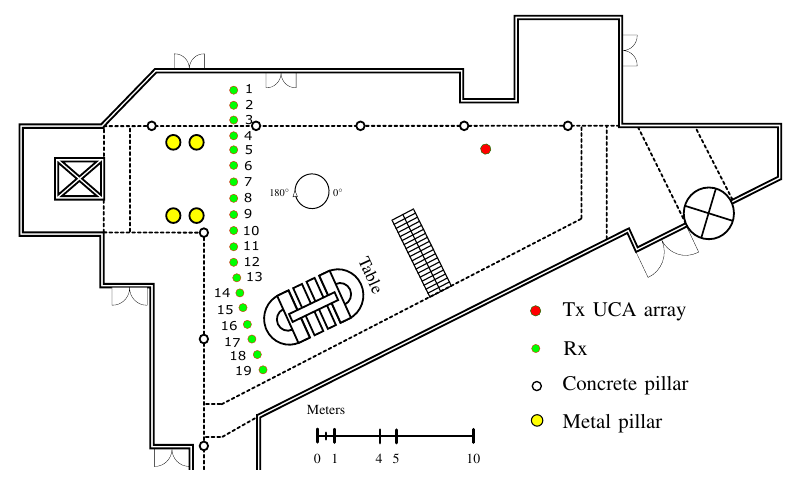}\label{fig:hall_scenario}}
 \subfigure[]{
\includegraphics[width=0.45\textwidth,trim=0 1.5cm 1cm 0, clip]{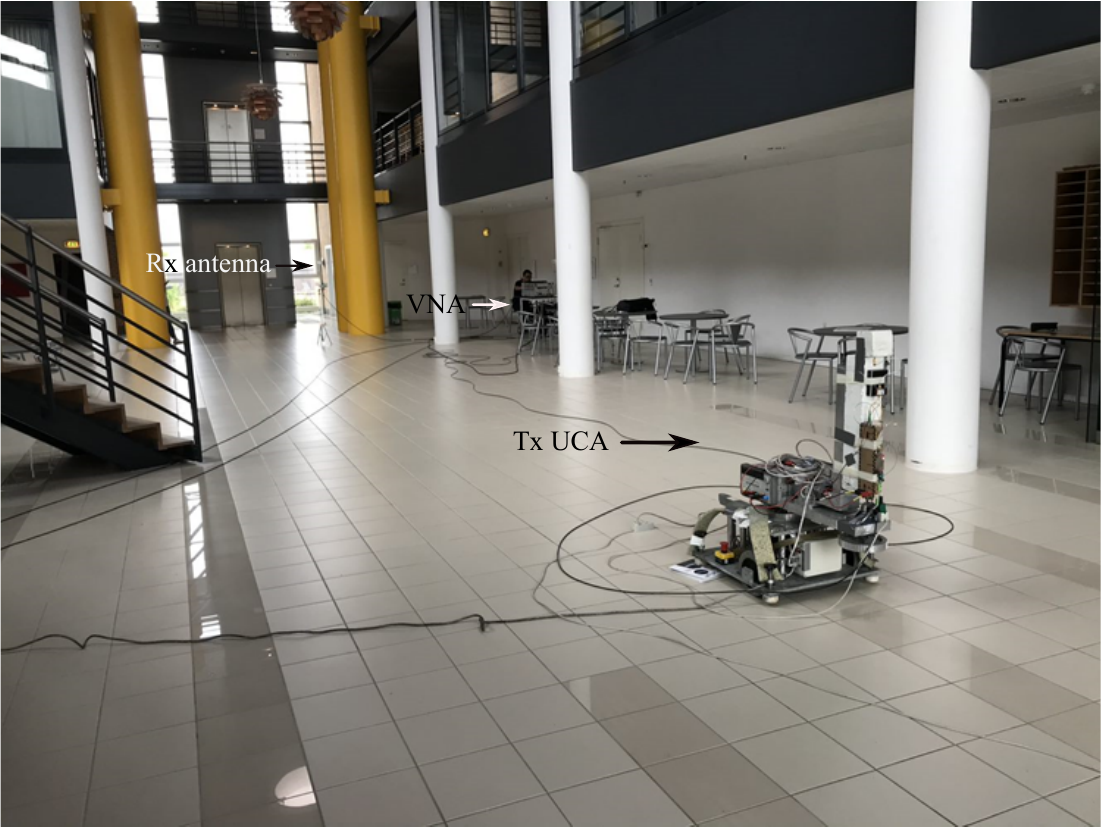}\label{fig:hall_scenario_photo}}
  \caption{ The layout (a) and a photo (b) of the indoor hall taken during the measurement campaign conducted at Aalborg University. The measurements corresponding to the $1^\textup{st}$ receive antenna array position are presented in this paper.}
      
\end{figure}

Measurements were conducted in a large hall scenario as illustrated in Fig.\,\ref{fig:hall_scenario}. A photo taken during the measurement campaign is also shown in Fig.\,\ref{fig:hall_scenario_photo}. The hall had a floor area of $44 \times 25~\text{m}^2$ with a height of approximately 10\,m. As shown in the picture and the layout sketch, tables, metallic pillars, concrete pillars, stairs, etc. were in the hall. The VNA measurements were taken with the ultra-wideband radio-over-fiber channel sounder developed at AAU \cite{8901446}. Quasi-omnidirectional biconical antennas \cite{8713575} were used. The receive antenna was fixed with a height of 1\,m to the ground while  transmit antenna was installed on a rotator and rotated with 720 uniform steps on a circle with a radius of  0.54\,m. In each step, the channel transfer function from 28~GHz to 30~GHz was swept with 1500 samples in the frequency domain. In this paper we analyse the first of the 19 different locations recorded. For this location, the transmitter-receiver distance was around 15\,m in NLOS condition.
\subsection{AAU Data, Outdoor Scenario}
\begin{figure}
\centering
\subfigure[]{\includegraphics[width=0.5\textwidth]{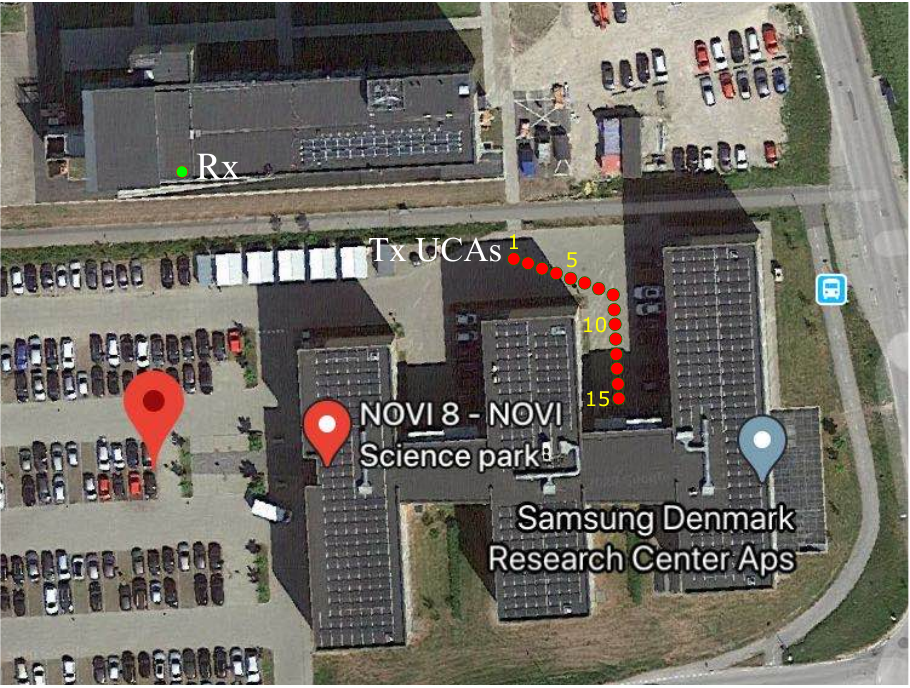}}
\subfigure[]{\includegraphics[width=0.5\textwidth]{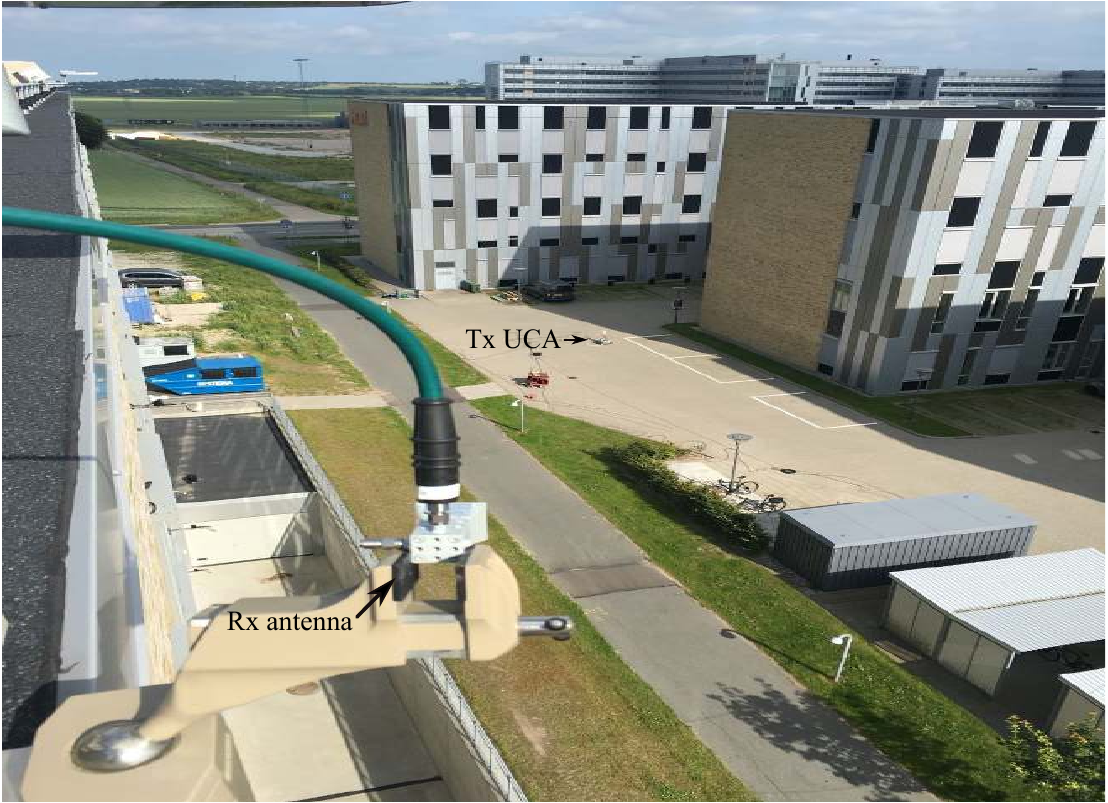}}
    \caption{Environment for the outdoor measurement campaign conducted at Aalborg University. Measurements from transmit antenna location number 7 are presented in this paper.}
    \label{fig:outdoor_scenario}
\end{figure}
\begin{table*}[]
	\centering
	\caption{
	Parameter estimates obtained using maximum likelihood estimation. Each entry corresponds to the estimate for some scalar parameter $\theta$, which corresponds to an element of either the 3-dimensional mean (column) vector $\boldsymbol{\mu}$ or the $3 \times 3$ dimensional covariance matrix $\boldsymbol{\Sigma}$. The number in bracket ($\delta$) is the half-width of the 95\% confidence interval for that parameter, so that the interval takes the form $(\theta-\delta,\theta +\delta)$.}
	    \label{tab:estimates}
\resizebox{0.7\textwidth}{!}{
\begin{tabular}{@{}ccccc@{}}
\toprule
                                      \textbf{Data set} & \textbf{Mean vector} $\hat{\boldsymbol{\mu}}(\pm \delta)$  &  \multicolumn{3}{c}{\textbf{Upper triangle of Covariance matrix} $\hat{\boldsymbol{\Sigma}} (\pm \delta)$}                 \\ \cmidrule(lr){1-2}\cmidrule(lr){3-5}
                                       
                                       
\multirow{3}{*}{\textbf{Lund}}    & --39 $(4\times 10^{-3})$                   & 2.8 (0.3)$\times 10^{-3}$ & 2.5 (0.3)$\times 10^{-3}$ & 1.4 (0.3)$\times 10^{-3}$ \\
                                      & --57 (4$\times 10^{-3}$)                   
                                      & 
                                      & 2.6 (0.3)$\times 10^{-3}$ & 2.1 (0.3)$\times 10^{-3}$ \\
                                      & --74 (6$\times 10^{-3}$)                   
                                      & 
                                      & 
                                      & 5.3 (0.6)$\times 10^{-3}$
                                      \\ \cmidrule(l){1-2}\cmidrule(l){3-5}
\multirow{3}{*}{\textbf{Lille}}   & --29 (0.03)                   & 0.19 (0.02)               & 0.15 (0.02)               & 0.11 (0.03)               \\
                                      & --47 (0.03)                   
                                      & 
                                      & 0.14 (0.01)
                                      & 0.19 (0.03)                \\
                                      & --63 (0.06)
                                      & 
                                      &
                                      & 0.70 (0.07)                \\ \cmidrule(l){1-2}\cmidrule(l){3-5}
\multirow{3}{*}{\textbf{AAU-Industry}} & --36 (0.31)                    & 2.34 (0.67)                & 1.36 (0.40)                & 1.24 (0.38)                \\
                                      & --53 (0.18)                 
                                      & 
                                      & 0.82 (0.23)    
                                      & 0.77 (0.23)                \\
                                      & --70 (0.18)    
                                      & 
                                      & 
                                      & 0.84 (0.24)               \\ \cmidrule(l){1-2}\cmidrule(l){3-5}
\multirow{3}{*}{\textbf{AAU-Hall}}     & --39 $(9 \times 10^{-3})$                   & 1.4 (0.14)$\times 10^{-2}$ & 1.2 (0.12)$\times 10^{-2}$ & 6.6 (0.76)$\times 10^{-3}$ \\
                                      & --56 $(7 \times 10^{-3})$ 
                                      & 
                                      & 1.0 (0.11)$\times 10^{-2}$ 
                                      & 6.2 (0.68)$\times 10^{-3}$ \\
                                      & --72 $(5 \times 10^{-3})$ 
                                      & 
                                      & 
                                      & 4.6 (0.48)$\times 10^{-3}$ \\ \cmidrule(l){1-2}\cmidrule(l){3-5}
\multirow{3}{*}{\textbf{AAU-Outdoor}}  & --40 $(1.2 \times 10^{-2})$                   & 1.3 (0.20)$\times 10^{-2}$ & 9.9 (0.14)$\times 10^{-3}$ & 5.2 (0.82)$\times 10^{-3}$ \\
                                      & --56 $(9 \times 10^{-3})$        
                                      & 
                                      & 7.6 (1.1)$\times 10^{-3}$
                                      & 4.2 (0.64)$\times 10^{-3}$ \\
                                      & --71 $(5 \times 10^{-3})$
                                      & 
                                      &
                                      & 2.7 (0.40)$\times 10^{-3}$ \\ \bottomrule
\end{tabular}
}
\end{table*}
\begin{table*}[]
	\centering
	\caption{AIC values for different model choices for the raw temporal moments. Best model is indicated in bold. Note that the joint AIC for the independent models is the sum of the AIC values of the three marginals.}
	\label{tab:aic}
\resizebox{0.8\textwidth}{!}{%
\begin{tabular}{@{}cccccc@{}}
\toprule
 \textbf{Data set}            & \textbf{\begin{tabular}[c]{@{}c@{}}Multivariate\\ Log-normal\end{tabular}} & \textbf{\begin{tabular}[c]{@{}c@{}}Multivariate\\ Gaussian\end{tabular}} & \textbf{\begin{tabular}[c]{@{}c@{}}Independent\\ Log-normal marginals\end{tabular}} &
             \textbf{\begin{tabular}[c]{@{}c@{}}Independent\\ Gaussian marginals\end{tabular}} &
             \textbf{\begin{tabular}[c]{@{}c@{}}Independent\\ Gamma marginals\end{tabular}} \\ \midrule
\textbf{Lund}    & \textbf{--219636.0}                                                         & --219573.9                                                                & --217787.6                     & --217750.2
            & --217777.0\\
\textbf{Lille}   & --208357.2                                                                  & \textbf{--208665.6}                                                       & --205657.4         & --204816.6
                & --205589.4\\
\textbf{AAU-Industry} & \textbf{--29815.61}                                                         & --28922.53                                                                & --29337.3                    & --28604.83
            & --29201.75\\
\textbf{AAU-Hall}     & \textbf{--247225.8}                                                         & --247212.2                                                                & --243329.2                     & --243348.8
            &--243342.9 \\
\textbf{AAU-Outdoor}  & --125244.8                                                                  & \textbf{--125286.7}                                                       & --122385.1     & --122342.4
            & --122374.1 \\ \bottomrule
\end{tabular}
}
\end{table*}
Outdoor measurements were  conducted in an open area to in-between the two buildings as shown in Fig.\,\ref{fig:outdoor_scenario}. The same channel sounder is used as in the indoor hall scenario. In this case, the transmitter antenna was rotated with a radius of  0.25\,m  in 360 uniform steps. In each step, the channel transfer function from 28~GHz to 30~GHz was swept with 2000 samples. The receive antenna was fixed on a roof edge with a height of around 20\,m. To increase the SNR, the receive antenna was replaced by a horn antenna with half-power-beamwidths around 30$^\circ$ in both azimuth and elevation. Moreover, its main beam was down tilted to appropriately  cover the transmit antenna.  Data was collected from 15 transmit transmit antenna locations as indicated in Fig.\,\ref{fig:outdoor_scenario}.  The data presented in this paper is from the 7$^\textup{th}$ location which was in LOS condition.
\section{Model Comparison}
To characterize the raw temporal moments jointly, their marginal distributions as well as their correlation structure needs to be well represented. We compare the proposed model against competing model choices for the available data-sets. 
\subsection{Model Comparison using AIC}
We compare the proposed joint model with the model of a multivariate Gaussian distribution. We also include three independent models for the raw temporal moments based on log-normal, Gaussian, and Gamma distributions. We omit comparison with the multivariate Gamma distributions in \cite{Kotz2000} as they did not give useful results when fitted to the raw temporal moments. Model comparison is done by computing the Akaike Information Criterion (AIC) value \cite{Akaike1974} of the competing models. AIC is a common tool for model selection that estimates the quality of different models relative to each other. It compares models through their likelihoods, but penalises models with a larger number of parameters $\kappa$. One motivation for this penalty comes from Ockham's razor, which states that, when comparing models, one should prefer the simplest model which explains the data well. The criterion is computed as follows
\begin{equation}
    \text{AIC} = -2\mathcal{L} + 2\kappa,
\end{equation}
where $\mathcal{L}$ is the maximized log-likelihood of the data. Given a set of models fitted by maximum likelihood to the same data, the preferred model is the one with the lowest AIC value. The reader is referred to \cite[Ch. 2]{Claeskens2008} for a detailed discussion. We also considered the Bayesian Information Criterion (BIC), which penalises more than AIC for a large number of parameters; see \cite{Ding2018} and \cite[Ch. 3]{Claeskens2008}. However, the ordering of the models was found to be the same for both the criteria, and therefore we omit the BIC values here.

The models are fitted to the five aforementioned data-sets by maximizing their likelihood. The parameter estimates obtained for the proposed model are reported in Tab.~\ref{tab:estimates}. The AIC values of the joint fit of the raw temporal moments are reported in Tab.~\ref{tab:aic}, with $\kappa = 9$ for the multivariate distributions, and $\kappa = 6$ for the independent marginal models. The proposed model comes out as the better choice for the joint fit for three out of five data-sets, with the multivariate Gaussian performing better for Lille Data and AAU-Outdoor. However, the AIC values for both the joint models are close to each other. It is evident that modeling the random variables independently leads to a significantly poorer fit than either of the joint models, no matter which distribution is chosen. We remark that using more complicated models such as copulas \cite{Nelsen2007} to model the dependency structure may lead to a better fit, but could be harder to interpret. 
\subsection{Log-normal vs. Gaussian Marginals}
\begin{figure*}
	 \centering
		 \includegraphics[width = 0.495\textwidth]{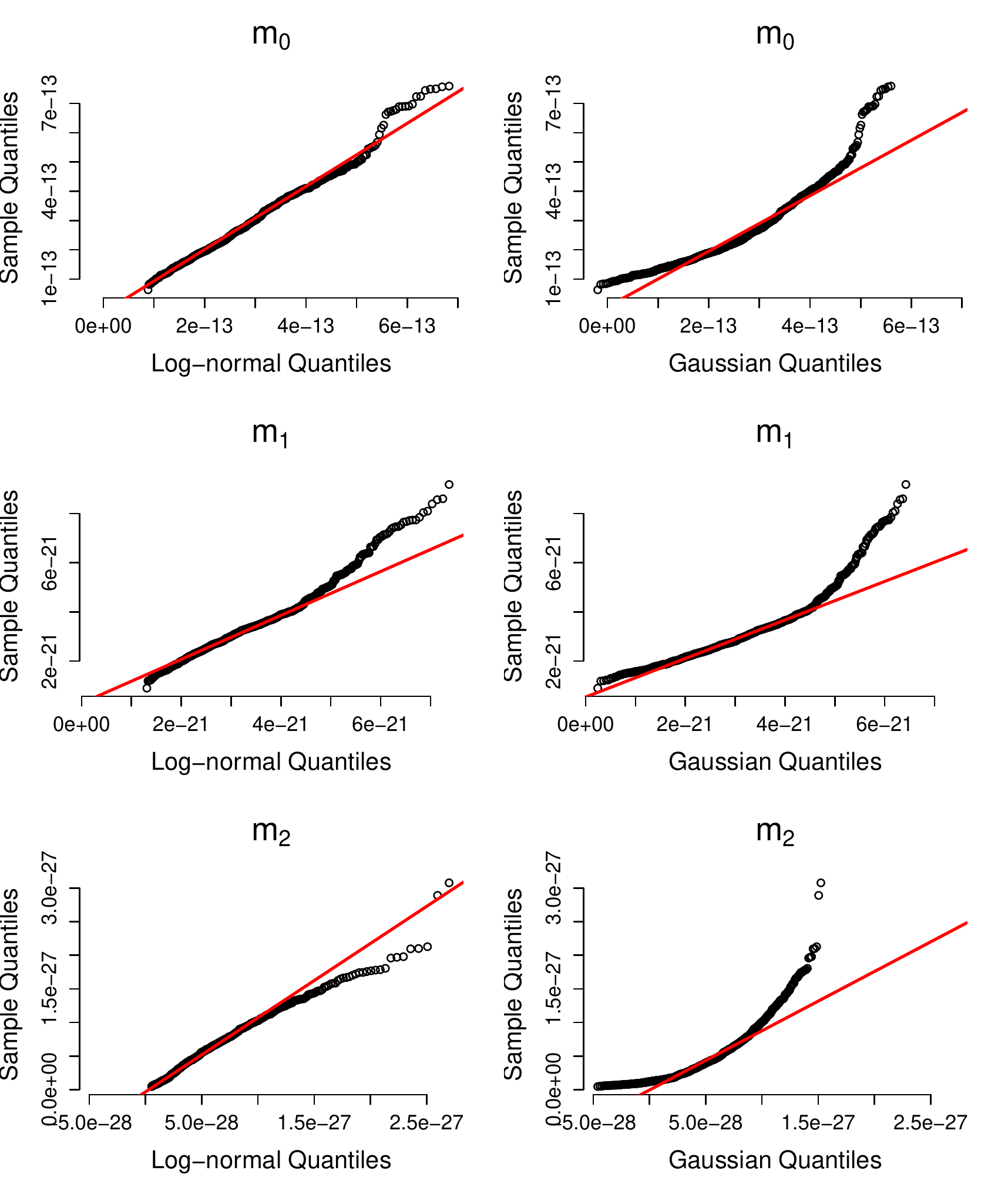}
		 \includegraphics[width = 0.495\textwidth]{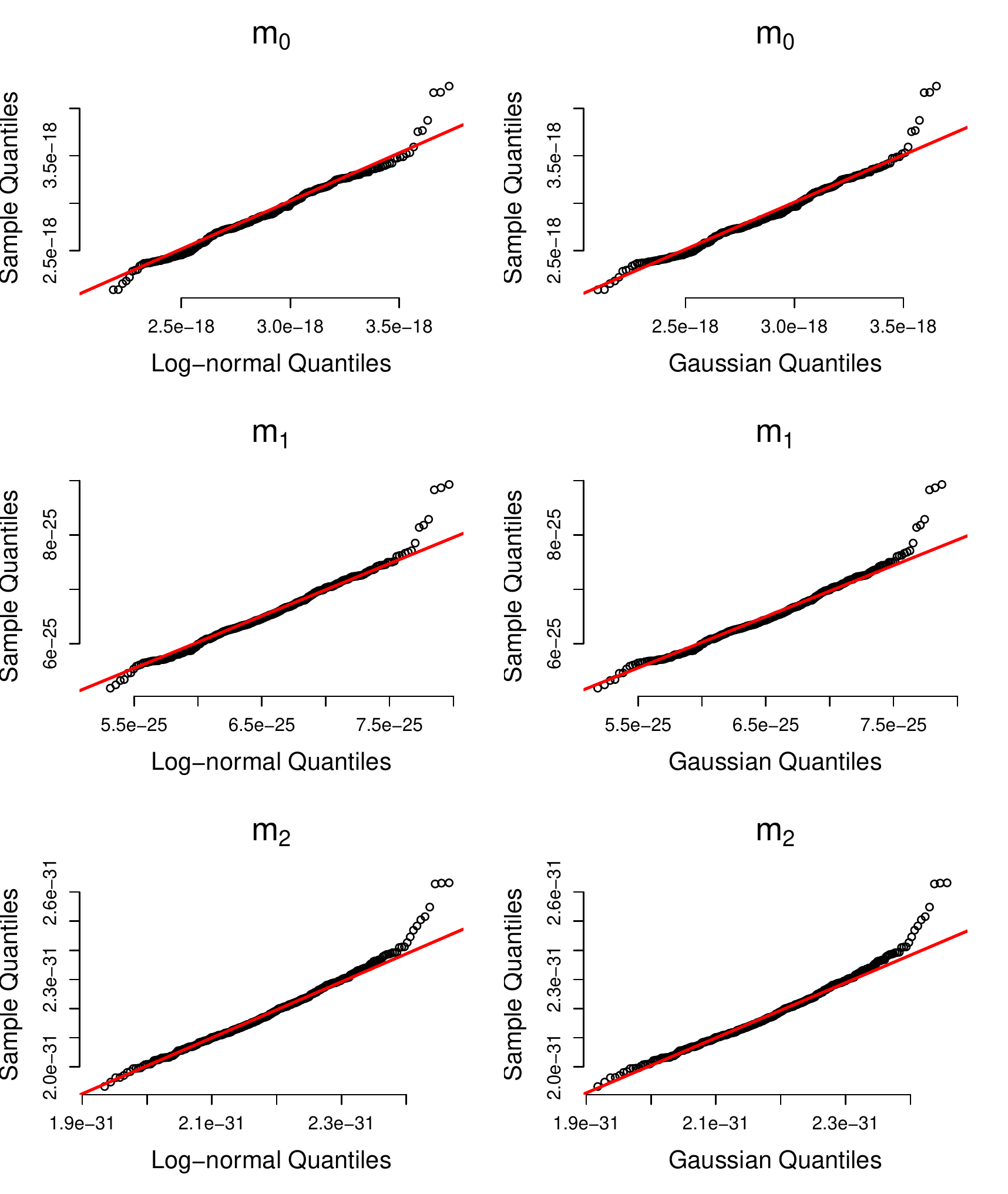}
	 \caption{Quantiles of the measured raw temporal moments from Lille (left) and AAU-Outdoor (right) data versus the theoretical quantiles of fitted log-normal and Gaussian distributions. The theoretical quantile-quantile line passing through the first and third quantile is shown in red.}
	 \label{fig:marginalComparison}
 \end{figure*}

We now compare the marginal fits of the multivariate log-normal and Gaussian distributions for modeling the raw temporal moments. To assess model fit, the quantiles of the data are plotted against the theoretical quantiles of the model being assessed. If the model is a good fit, then the quantiles of the data and the theoretical quantiles should be close to one another, and the points will hence lie approximately on a straight line. On the other hand, any deviation from this line might indicate issues with the fit of the model. See e.g. \cite[Sec. 10.2]{Rice1994} for more details. We show the Q-Q plots for two of the five data-sets, namely the Lille and the AAU-Outdoor data, in Fig.~\ref{fig:marginalComparison}, as they highlight the difference between the fits obtained from both the distributions. The Q-Q plots of AAU-Outdoor data is representative of what we observed for the other data-sets, therefore we exclude reporting them.


We observe that for AAU-Outdoor data, the marginals are well-modeled by both the log-normal and the Gaussian distributions. The fit is similar for both distributions, and it is not apparent which model performs better. As can be seen in Fig.~\ref{fig:rawFit}, the marginals in AAU-Outdoor data are very close to being symmetric, which means that the Gaussian fits well. However, for the Lille data, it is evident that the log-normal distribution outperforms the Gaussian in terms of the marginals. The log-normal is able to model the left tail and the center of the distribution very well, but sometimes performs poorly for the right tail. On the other hand, the Gaussian is not able to model either of the tails. Moreover, the Gaussian assigns non-zero probabilities to quantiles below zero, which is not the case for the data as temporal moments are non-negative random variables. Hence, the multivariate log-normal is a better choice. Note that a good marginal fit does not imply good overall fit in terms of AIC and vice-versa, as is the case for Lille data. This is simply because the AIC measures a different property of the model which does not require the marginals to fit perfectly.

The deviation of the right tail of the data from the fitted marginals is to be expected due to the low number of such extreme points. Such points are in-frequent and could potentially arise due to a number of factors such as noise, interference, measurement conditions, etc. Therefore, we argue that the right tail is not as important to model perfectly, and thus make no adjustment for it. However, this should be scrutinized further in applications where this effect could be important.
\section{Model Fit to Raw Temporal Moments}
\begin{figure*}
	 \centering
		 \includegraphics[width = 0.48\textwidth]{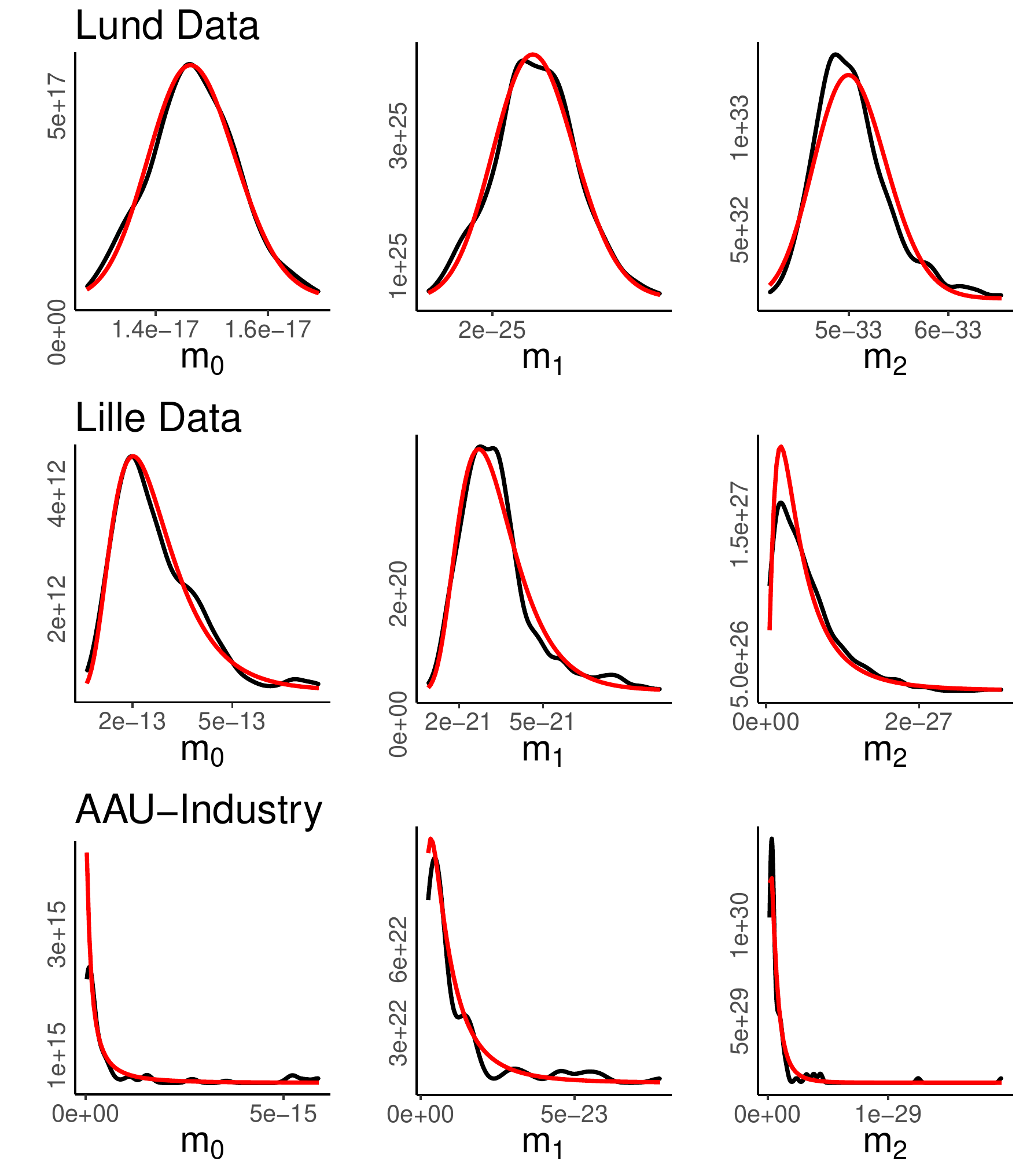}
		 \includegraphics[width = 0.48\textwidth]{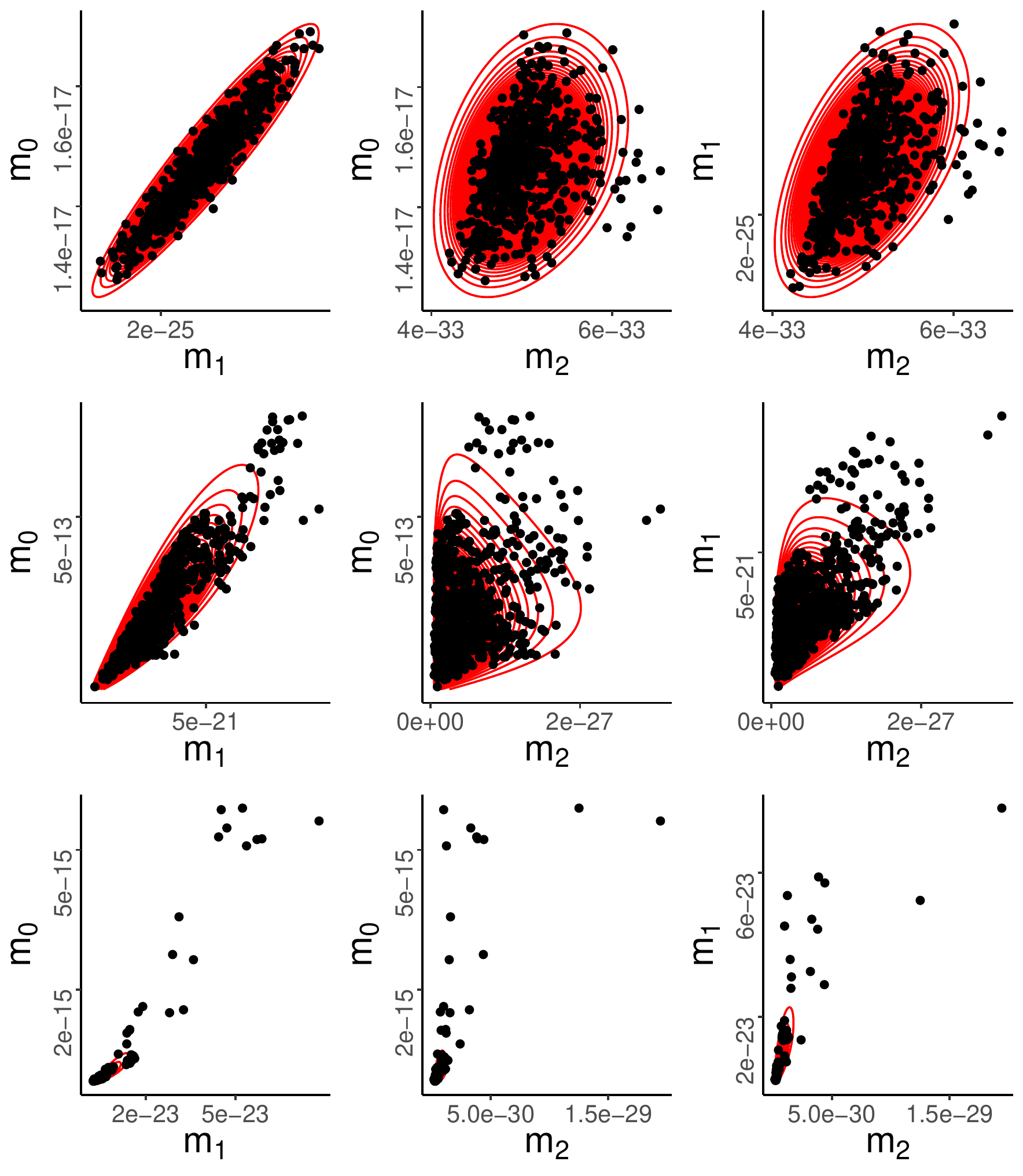}
		 \includegraphics[width = 0.48\textwidth]{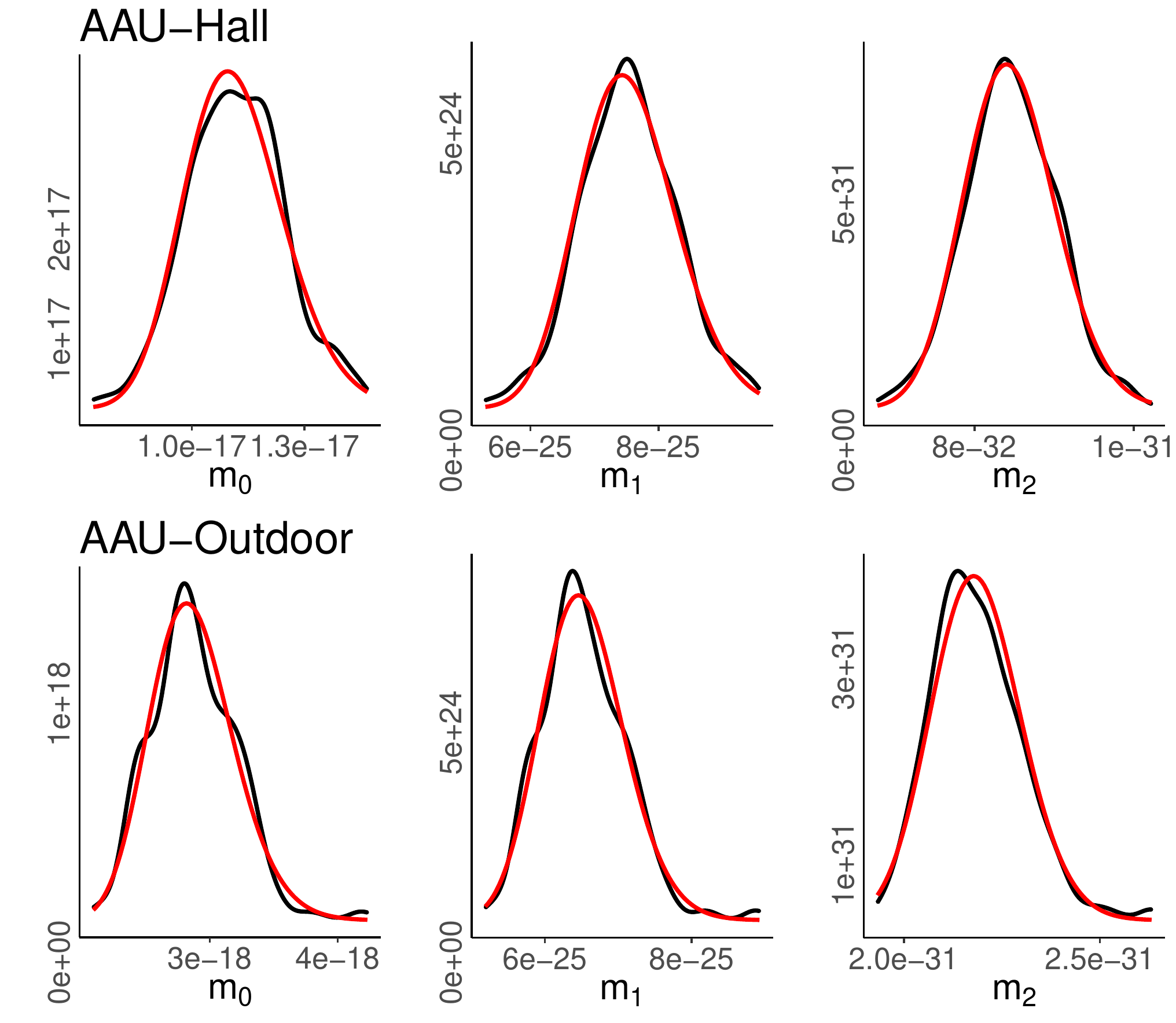}
		 \includegraphics[width = 0.48\textwidth]{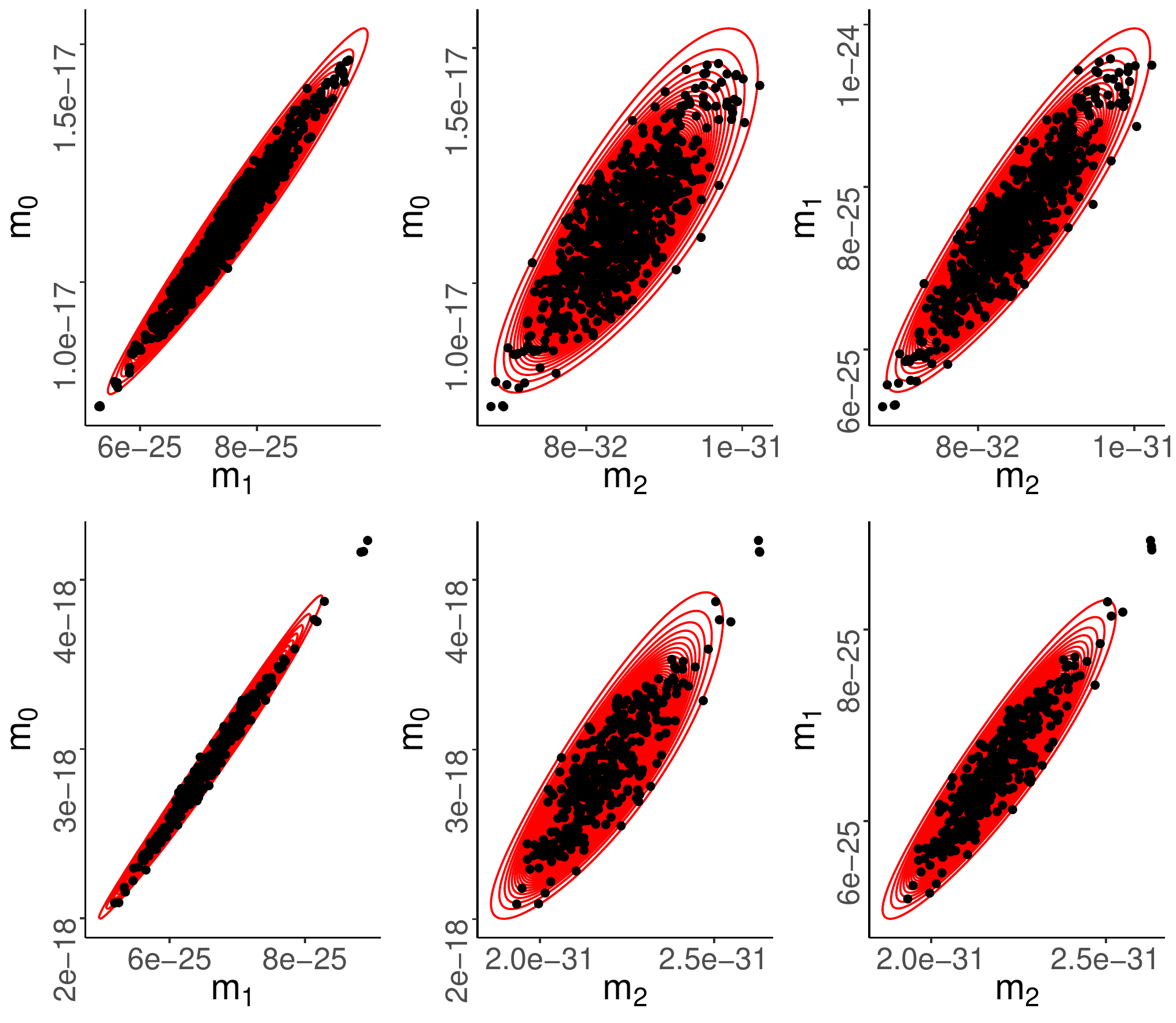}
		 \caption{Density estimates and scatter plots of raw temporal moments obtained from the different measurements (shown in black) versus the density and contour plots of the fitted proposed model (shown in red). Each row corresponds to one of the data-sets. All the axes are in linear scale. The parameter estimates are in Tab.~\ref{tab:estimates}.}
	 \label{fig:rawFit}
 \end{figure*}

The parameter estimates, obtained by fitting the proposed model to the data-sets using \eqref{eq:Muestimates} and \eqref{eq:Sigmaestimates}, are reported in Tab.~\ref{tab:estimates}. We also compute and report the 95\% confidence intervals for each of the estimates in Tab.~\ref{tab:estimates}, see Appendix~\ref{app:confInterval} for details. The confidence intervals are very small for the mean estimates, and an order of magnitude smaller for the covariance estimates. The fit of the proposed model to the various data-sets is shown in Fig.~\ref{fig:rawFit} where each row corresponds to a particular data-set. The marginal distributions of the data and the fitted model is shown on the left, while 2D scatter plots for all pairs of temporal moments are shown on the right along with contour lines of the fitted distribution.

Firstly, we observe in Fig.~\ref{fig:rawFit} that the distribution of the raw temporal moments varies across the different data-sets. This is attributed to the contrasting scenarios that the measurements were taken in, along with the use of different equipment, antennas, and measurement settings. We also observe that the raw temporal moments are highly correlated random variables. Marginal distributions for Lille and AAU-Industry data are skewed, while those from other data-sets are more symmetric. We notice a fanning out of the scatter plots on the top-right of all the indoor data-sets, which is not present in the outdoor data. Despite the variability in the data, the proposed model fits the data well, even the skewed ones. There is a high correlation between the raw moments, in particular between $m_0$ and $m_1$, since the basis functions used to compute them in \eqref{eq:temporalMoments} are not orthogonal. This is captured well by the model.
\section{Model Fit to Standardized Moments}
\begin{figure*}
	 \centering
		 \includegraphics[width = 0.48\textwidth]{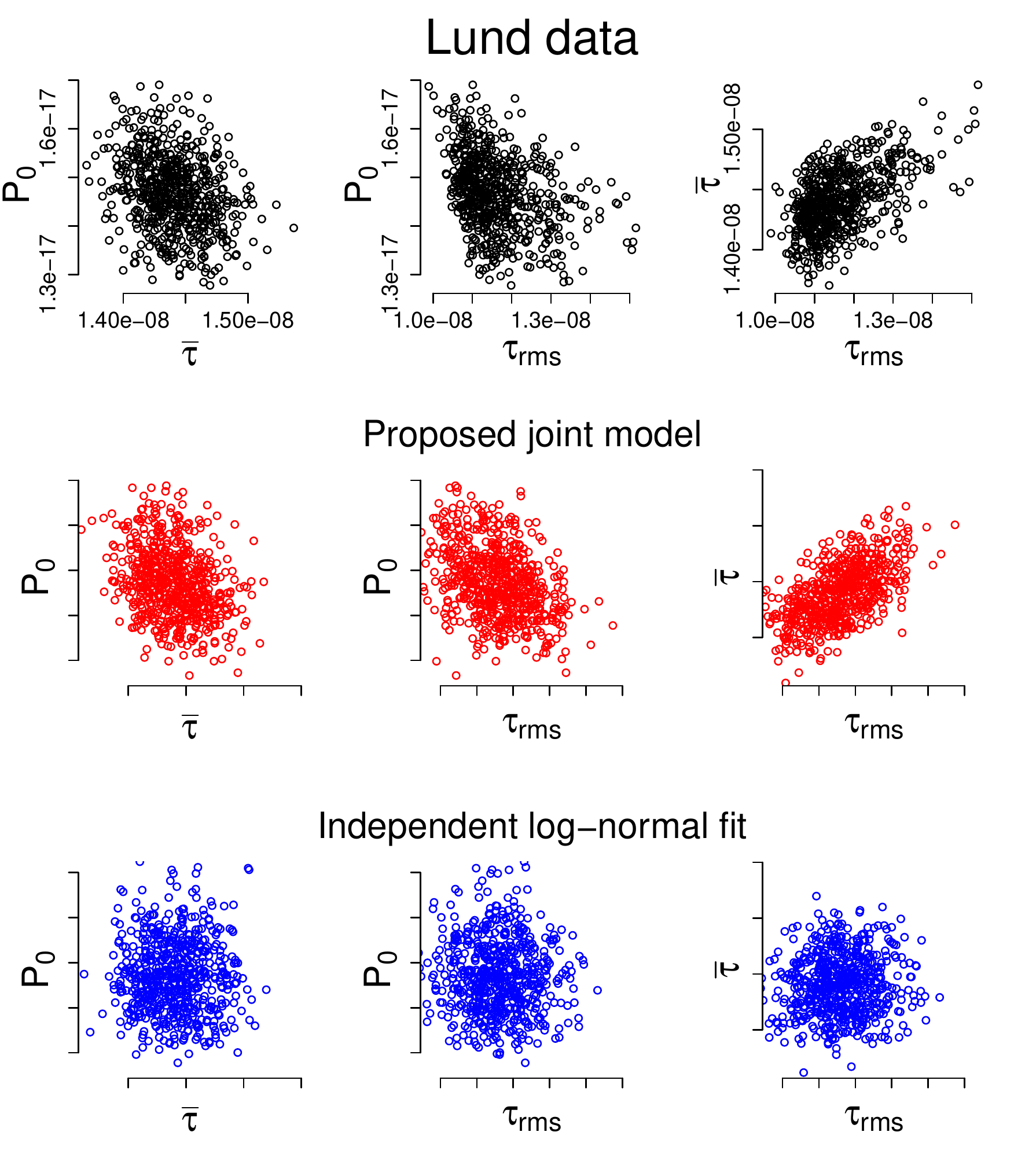}
		 \includegraphics[width = 0.48\textwidth]{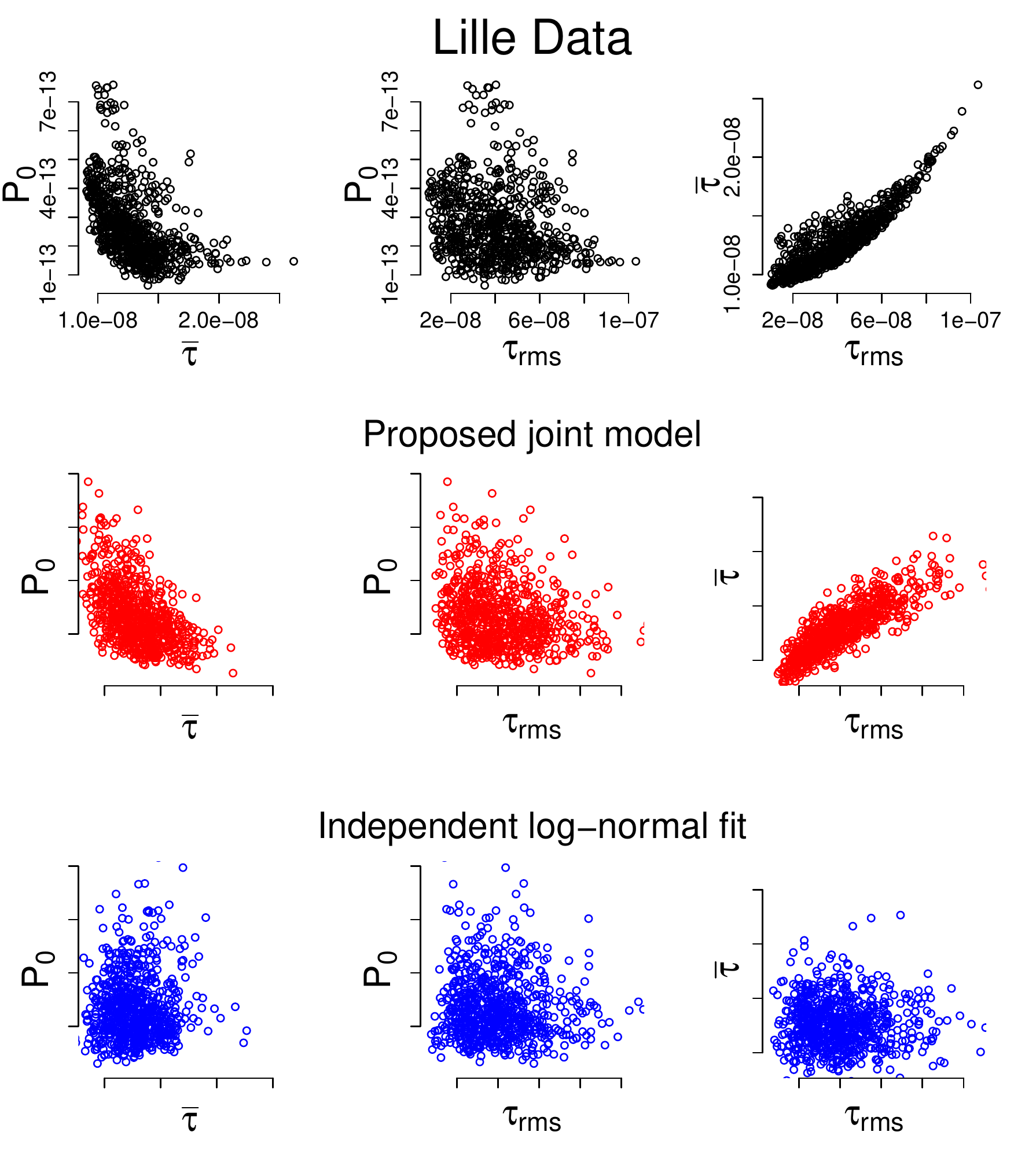}
		 \includegraphics[width = 0.48\textwidth]{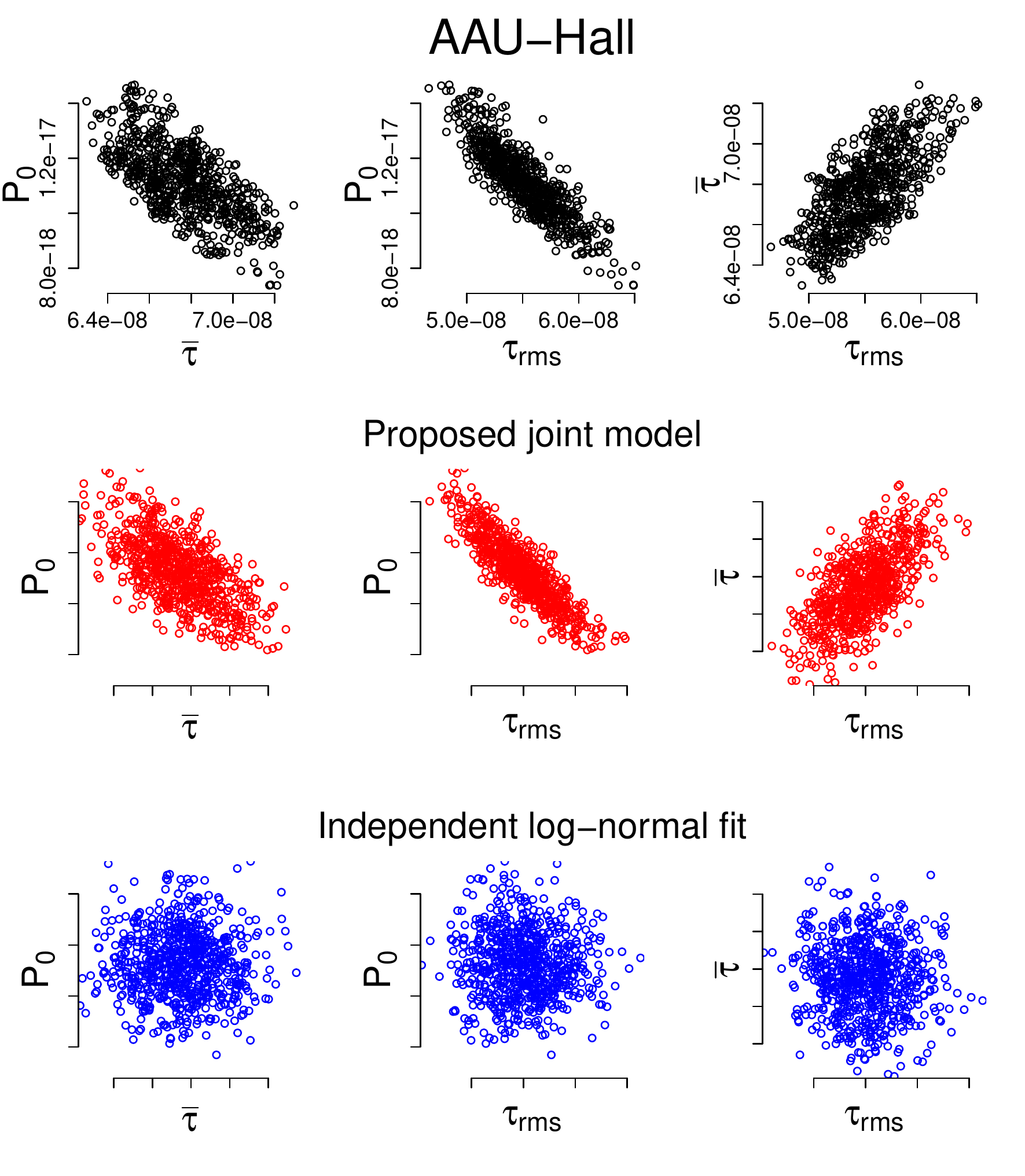}
		 \includegraphics[width = 0.48\textwidth]{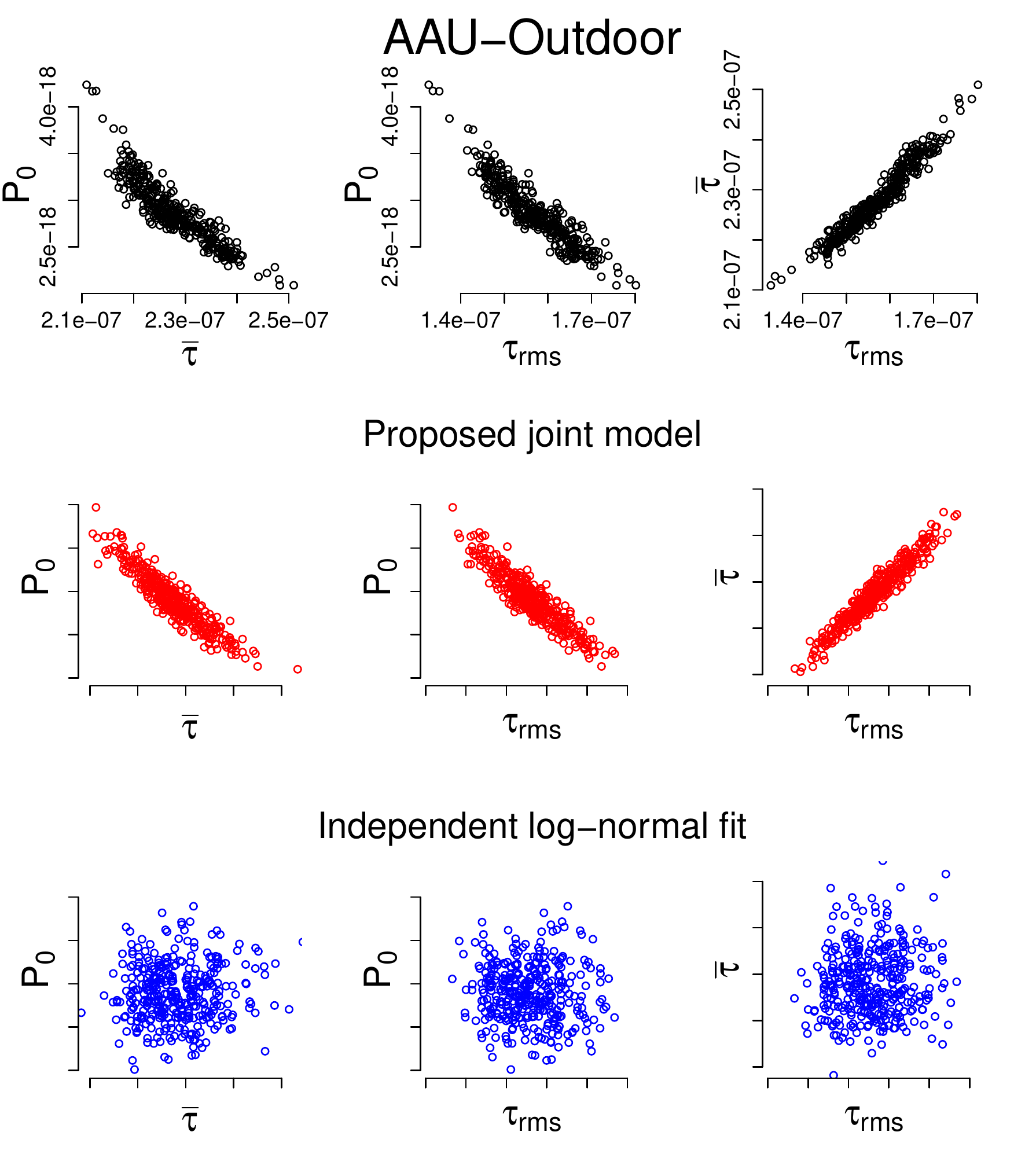}
	 \caption{Scatter plots of received power, mean delay, and rms delay spread from data (in black), and from the proposed model (in red). The samples simulated by independently fitting log-normal marginals to $P_0$, $\bar \tau$, and $\tau_{\mathrm{rms}}$ from the data are in blue. Number of points simulated is same as in the measurements. The scales of the corresponding plots are the same.}
	 \label{fig:standardizedFit}
 \end{figure*}

We now compare the distribution of the standardized temporal moments obtained from the measurements with those from the proposed model. Mean delay and rms delay spread are computed from the raw temporal moments using \eqref{eq:standardizedMoments}, while the received power is equal to $m_0$. Pair-wise scatter plots of $P_0$, $\bar \tau$, and $\tau_{\mathrm{rms}}$ from the data and the proposed joint model are shown in Fig.~\ref{fig:standardizedFit}. We also include the samples obtained from independently fitting a log-normal distribution to the standardized moments from the data-sets. The log-normal is chosen as it was the best in terms of AIC amongst the independent models as per Tab.~\ref{tab:aic}. Here we exclude the AAU-Industry data as the low number of sample points makes it difficult to make any useful conclusions on the correlation behavior. We observe in Fig.~\ref{fig:standardizedFit} that the standardized temporal moments are also dependent random variables, and the proposed model is able to capture their dependency structure. In contrast, correlation information between the variables is lost when they are simulated independently.

Sample Pearson correlation coefficients between $P_0$, $\bar \tau$, and $\tau_{\mathrm{rms}}$ from the data are given in Tab.~\ref{tab:correlation}. For paired samples $\{(a_1,b_1), \dots, (a_m,b_m)\}$, the sample Pearson correlation coefficient is defined as 
\begin{equation}
	\hat{\rho}_{a,b} = \frac{\sum_{j=1}^m (a_j - \bar{a}) (b_j - \bar{b})}{\sqrt{\sum_{j=1}^m (a_j - \bar{a})^2} \sqrt{\sum_{j=1}^m (b_j - \bar{b})^2}},
\end{equation}
where $\bar{a}$ and $\bar{b}$ are the sample means. We also compute 95\% confidence intervals for the correlation estimates using the bootstrap method \cite[Chapter~6]{Efron1994}. The correlation coefficients obtained from the fitted model, computed from 10,000 samples to get a robust estimate, are also reported in Tab.~\ref{tab:correlation}. Mean delay and rms delay spread have a positive correlation that varies from 0.53 for the Lund data to as high as 0.97 for AAU-Outdoor. The received power is negatively correlated with both $\tau$ and $\tau_{\mathrm{rms}}$. In general, the correlation tends to increase with the size of the environment, with the outdoor case being highly correlated. The model is able to replicate the varying correlation between $P_0$ and $\tau_{\mathrm{rms}}$ that is observed in the data, as opposed to having a fixed correlation coefficient suggested in \cite{Greenstein1997}. Note that the correlation coefficient between $\bar\tau$ and $\tau_{\mathrm{rms}}$ for the model fitted to the Lille data-set is not within the bootstrap interval. This is due to the banana-like shape of their scatter plot which is not replicated by the model, see Fig.~\ref{fig:standardizedFit}.
\begin{table*}[]
	\centering
	\caption{Sample Pearson correlation coefficients between standardized temporal moments of measured data. The correlation coefficients for the model is computed using 10,000 samples of simulated data. The number in parenthesis ($\epsilon$) is the 95\% bootstrap confidence interval of the correlation estimates computed using 1000 resamples, such that the interval is of the form $(\rho - \epsilon, \rho + \epsilon)$.}
	\label{tab:correlation}
\resizebox{0.75\textwidth}{!}{
\begin{tabular}{@{}ccccccc@{}}
\toprule
  \textbf{Data set}                   & \multicolumn{2}{c}{$ \hat{\rho}_{P_0, \bar{\tau}} $} & \multicolumn{2}{c}{$ \hat{\rho}_{P_0, \tau_{\mathrm{rms}}} $} & \multicolumn{2}{c}{$\hat{\rho}_{\bar{\tau}, \tau_{\mathrm{rms}}} $} \\
                     & \textbf{Data}            & \textbf{Model}            & \textbf{Data}                 & \textbf{Model}                & \textbf{Data}                    & \textbf{Model}                   \\ \cmidrule(l){1-3}\cmidrule(l){4-5}\cmidrule(l){6-7}
\textbf{Lund}   & --0.28 ($\pm$ 0.06)            & --0.28                    & --0.35 ($\pm$0.05)                 & --0.36                        & 0.53 ($\pm$0.05)                      & 0.52                             \\
\textbf{Lille}  & --0.48 ($\pm$0.03)            & --0.51                    & --0.20 ($\pm$0.05)                 & --0.19                        & 0.89 ($\pm$0.02)                      & 0.83                             \\
\textbf{AAU-Hall}    & --0.66 ($\pm$0.03)            & --0.65                    & --0.87 ($\pm$0.02)                 & --0.87                        & 0.70 ($\pm$0.03)                      & 0.70                             \\
\textbf{AAU-Outdoor} & --0.91 ($\pm$0.01)            & --0.92                    & --0.93 ($\pm$0.01)                 & --0.93                        & 0.97 ($\pm$0.004)                     & 0.97                             \\ \bottomrule
\end{tabular}
}
\end{table*}

\section{Conclusions}
Joint modeling of received power, mean delay, and rms delay spread provides more accurate models in a range of scenarios as opposed to independent modeling. The proposed model of the multivariate log-normal distribution seems to be a reasonable choice for simulating these standardized moments, however the fit can be improved by using more complex models. The proposed model is simple, easy to simulate from, and easy to fit to new measurements in both indoor and outdoor settings using standard estimators. The raw temporal moments are dependent random variables which should be simulated jointly; as a result, the same is also true for the standardized temporal moments. The correlation of these moments changes from scenario to scenario, but can be inferred efficiently in each case.

In the light of the strong correlation observed in the measurements, assuming independence might lead to significant errors in some applications. Hence, reporting of the marginal distributions of the standardized moments is insufficient and a clearer picture can be obtained by considering both their means and covariances. The correlation between these standardized moments can be used to validate multipath models instead of just their marginal fits. The correlation should also be accounted for in the analysis and simulation of radio channels.


The means and covariances of the temporal moments potentially depend on a number of physical factors. The relation between the means and the transmitter-receiver distance  has been studied for indoor scenarios. However, the effect of the distance on the covariance matrix is presently unclear. For multipath models, the covariance matrix is known to depend on the first- and second-order intensity functions which governs the arrival process. Since both intensity functions are affected by antenna directivity, the covariance matrix should also be. Nevertheless, these effects are not yet well-understood and should be the topic of further studies.
\section*{Acknowledgment}
The authors would like to thank Dr. Carl Gustafson and Prof. Fredrik Tufvesson (Lund University) for providing the measurement data. This work is supported by the Danish Council for Independent Research, grant no. DFF 7017-00265 and performed within the framework of the COST Action CA15104 IRACON. Ramoni Adeogun was supported by grant no. DFF 9041- 00146B. Fran\c{c}ois-Xavier Briol was supported by the Lloyds Register Foundation Programme on Data-Centric Engineering at The Alan Turing Institute under the EPSRC grant [EP/N510129/1].
%
\appendices
\section{Parameter Inference for the Log-Normal} \label{app:confInterval}
%


In this appendix, we recall how to derive the maximum likelihood estimates and related confidence intervals for a log-normal distribution.
Let $ Y = (Y_1, \dots, Y_d) $ be a multivariate log-normal random variable. We will denote this distribution $ \mathcal{LN}(\boldsymbol \mu, \boldsymbol  \Sigma)$, where $\boldsymbol \mu$ and $\boldsymbol \Sigma$ denote the parameters. Then, $X= (X_1, \dots, X_d) = (\log(Y_1),\ldots,\log(Y_d))$ is a multivariate Gaussian random variable with mean vector $\boldsymbol \mu$ and covariance matrix $\boldsymbol \Sigma$. Since the maximum likelihood estimator is invariant to one-to-one transformations of the data, we can simply take the logarithm of our data points and compute the maximum likelihood estimate corresponding to Gaussian data. Given $N$ iid observations $\{\mathbf{y}_i\}_{i=1}^N$, we hence compute $\mathbf{x}_i = \log \mathbf{y}_i$ for $i=1,\ldots,N$, and return the following estimates
\begin{align}
	\hat{\boldsymbol \mu} &= \frac{1}{N} \sum_{i=1}^{N}\mathbf{x}_i, \quad \text{and}\\
	\hat{\boldsymbol \Sigma}  &= \frac{1}{N} \sum_{i=1}^{N} \left( \mathbf{x}_i -  \hat{\boldsymbol{\mu}}\right) \left( \mathbf{x}_i -  \hat{\boldsymbol{\mu}}\right)^\top.
\end{align}

Now, let the $ K $ free parameters be combined into a single vector $ \boldsymbol \theta = (\boldsymbol \alpha, \boldsymbol \beta)$, where $\boldsymbol \alpha = (\mu_1, \dots, \mu_d)$, and $\boldsymbol \beta= (\Sigma_{11}, \dots, \Sigma_{dd}) $. Note that $ \Sigma_{ij} =  \Sigma_{ji}$. The Fisher information matrix reads
%
\begin{equation}
I( \boldsymbol \alpha, \boldsymbol \beta) = \begin{bmatrix}
I( \boldsymbol \alpha) 	& \mathbf{0}\\
\mathbf{0}				& I( \boldsymbol \beta)
\end{bmatrix}
\end{equation}
where, for $1\leq m,n \leq K$, the $(m,n)$ entry of the matrix is
\begin{align}
	I(\boldsymbol \alpha)_{m,n} &= \frac{\partial \boldsymbol \mu^\top}{\partial \alpha_m} \boldsymbol \Sigma^{-1} \frac{\partial \boldsymbol \mu}{\partial \alpha_n}, \quad 1\leq m,n \leq d\\
	I(\boldsymbol \beta)_{m,n} &= \frac{1}{2} \mathrm{tr}\left(\boldsymbol \Sigma^{-1} \frac{\partial \boldsymbol \Sigma}{\partial \beta_m} \boldsymbol \Sigma^{-1} \frac{\partial \boldsymbol \Sigma}{\partial \beta_n} \right).
\end{align}
On further simplification, the entries of the Fisher information matrix become
\begin{align}
 I(\boldsymbol \alpha)_{m,n} &=  \boldsymbol \Sigma_{mn}^{-1},\\
 I(\boldsymbol \beta)_{m,n} &= \frac{1}{2} \mathrm{tr}\left(\boldsymbol \Sigma^{-1} \mathbf{E}_{m}\boldsymbol \Sigma^{-1} \mathbf{E}_{n} \right),
\end{align}
where $ \mathbf{E}_{m} $ is a $ d\times d $ matrix of all zeros except the $(i,i)$ entry corresponding to $\boldsymbol \beta_m = \Sigma_{ii}$ which is 1. Note that for $\boldsymbol \beta_m = \Sigma_{ij}, i\neq j $, both $ (i,j) $ and $ (j,i) $ entry of $ \mathbf{E}_{m} $ will be 1. Same goes for $ \mathbf{E}_n $. The 95\% confidence interval for the $m^\textup{th}$ parameter of the Gaussian, $(\theta_m \pm \delta_m)$ is


$$\theta_m \pm  \frac{1.96}{\sqrt{N}}\sqrt{\mathbf{I}_{m,m}^{-1}}$$
where $\mathbf{I}_{m,m}^{-1}$ is the $(m,m)$ entry of $\mathbf{I}^{-1}$. 

\bibliography{bibliography.bib}

\begin{thebibliography}{10}
\providecommand{\url}[1]{#1}
\csname url@samestyle\endcsname
\providecommand{\newblock}{\relax}
\providecommand{\bibinfo}[2]{#2}
\providecommand{\BIBentrySTDinterwordspacing}{\spaceskip=0pt\relax}
\providecommand{\BIBentryALTinterwordstretchfactor}{4}
\providecommand{\BIBentryALTinterwordspacing}{\spaceskip=\fontdimen2\font plus
\BIBentryALTinterwordstretchfactor\fontdimen3\font minus
  \fontdimen4\font\relax}
\providecommand{\BIBforeignlanguage}[2]{{%
\expandafter\ifx\csname l@#1\endcsname\relax
\typeout{** WARNING: IEEEtran.bst: No hyphenation pattern has been}%
\typeout{** loaded for the language `#1'. Using the pattern for}%
\typeout{** the default language instead.}%
\else
\language=\csname l@#1\endcsname
\fi
#2}}
\providecommand{\BIBdecl}{\relax}
\BIBdecl

\bibitem{Goldsmith2005}
A.~Goldsmith, \emph{Wireless Communications}.\hskip 1em plus 0.5em minus
  0.4em\relax Cambridge University Press, Aug 2005.

\bibitem{Wu2008}
W.-D. Wu, C.-H. Wang, C.-C. Chao, and K.~Witrisal, ``On parameter estimation
  for ultra-wideband channels with clustering phenomenon,'' in \emph{2008
  {IEEE} 68th Veh. Technol. Conf.}\hskip 1em plus 0.5em minus 0.4em\relax
  {IEEE}, Sep 2008.

\bibitem{AyushURSI}
A.~Bharti, R.~Adeogun, and T.~Pedersen, ``{Parameter Estimation for Stochastic
  Channel Models using Temporal Moments},'' in \emph{Proc. 2019 IEEE Int. Symp.
  on Antennas and Propag. and USNC-URSI Radio Sci. Meeting}, 2019.

\bibitem{AyushSPAWC19}
------, ``{Estimator for Stochastic Channel Model without Multipath Extraction
  using Temporal Moments},'' in \emph{20th IEEE Int. Workshop on Signal
  Process. Advances in Wireless Commun. (SPAWC)}, 2019.

\bibitem{AyushABC}
A.~Bharti and T.~Pedersen, ``Calibration of stochastic channel models using
  approximate {B}ayesian computation,'' in \emph{Proc. IEEE Global Commun.
  Conf. Workshops}, 2019.

\bibitem{Bharti2020}
A.~Bharti, R.~Adeogun, and T.~Pedersen, ``Learning parameters of stochastic
  radio channel models from summaries,'' \emph{{IEEE} Open J. of Antennas and
  Propag.}, pp. 1--1, 2020.

\bibitem{RamoniMLAWPL}
R.~{Adeogun}, ``Calibration of stochastic radio propagation models using
  machine learning,'' \emph{IEEE Antennas and Wireless Propag. Letters},
  vol.~18, no.~12, pp. 2538--2542, Dec 2019.

\bibitem{Latinovic2020}
Z.~Latinovic and H.~Huang, ``A channel model for indoor time-of-arrival
  ranging,'' \emph{{IEEE} Trans. on Wireless Commun.}, vol.~19, no.~2, pp.
  1415--1428, feb 2020.

\bibitem{Awad2008}
M.~K. {Awad}, K.~T. {Wong}, and Z.~{Li}, ``An integrated overview of the open
  literature's empirical data on the indoor radiowave channel's delay
  properties,'' \emph{IEEE Trans. on Antennas and Propag.}, vol.~56, no.~5, pp.
  1451--1468, May 2008.

\bibitem{Cox1975}
D.~{Cox} and R.~{Leck}, ``Distributions of multipath delay spread and average
  excess delay for 910 {MHz} urban mobile radio paths,'' \emph{IEEE Trans. on
  Antennas and Propag.}, vol.~23, no.~2, pp. 206--213, March 1975.

\bibitem{Greenstein1997}
L.~Greenstein, V.~Erceg, Y.~Yeh, and M.~Clark, ``A new path-gain/delay-spread
  propagation model for digital cellular channels,'' \emph{IEEE Trans. Veh.
  Technol.}, vol.~46, no.~2, pp. 477--485, May 1997.

\bibitem{Fischer2013}
J.~Fischer, M.~Grossmann, W.~Felber, M.~Landmann, and A.~Heuberger, ``A novel
  delay spread distribution model for {VHF} and {UHF} mobile-to-mobile
  channels,'' in \emph{2013 7th Eur. Conf. on Antennas and Propag., EuCAP
  2013}, Apr 2013, pp. 469--472.

\bibitem{Wang2019}
G.~Wang, G.~Zhu, S.~Lin, J.~Ding, D.~Fei, and H.~Zhang, ``Channel measurement
  and modeling in highway scenario at 460 {MHz},'' in \emph{2019 {IEEE}
  International Symposium on Antennas and Propagation and {USNC}-{URSI} Radio
  Science Meeting}.\hskip 1em plus 0.5em minus 0.4em\relax {IEEE}, jul 2019.

\bibitem{Li2019}
J.~Li, B.~Ai, R.~He, M.~Yang, and Z.~Zhong, ``Multi-frequency channel
  characterization for massive {MIMO} communications in lobby environment,''
  \emph{China Commun.}, vol.~16, no.~9, pp. 79--92, sep 2019.

\bibitem{Zhang2019}
X.~Zhang, G.~Qiu, J.~Zhang, L.~Tian, P.~Tang, and T.~Jiang, ``Analysis of
  millimeter-wave channel characteristics based on channel measurements in
  indoor environments at 39 {GHz},'' in \emph{2019 11th International
  Conference on Wireless Communications and Signal Processing ({WCSP})}.\hskip
  1em plus 0.5em minus 0.4em\relax {IEEE}, oct 2019.

\bibitem{Yu2017}
Y.~{Yu}, Y.~{Liu}, W.~{Lu}, and H.~{Zhu}, ``Measurement and empirical modelling
  of root mean square delay spread in indoor femtocells scenarios,'' \emph{IET
  Commun.}, vol.~11, no.~13, pp. 2125--2131, 2017.

\bibitem{Schmieder2020}
M.~Schmieder, T.~Eichler, S.~Wittig, M.~Peter, and W.~Keusgen, ``Measurement
  and characterization of an indoor industrial environment at 3.7 and 28
  {GHz},'' in \emph{2020 14th European Conference on Antennas and Propagation
  ({EuCAP})}.\hskip 1em plus 0.5em minus 0.4em\relax {IEEE}, mar 2020.

\bibitem{Yu2020}
J.~Yu, W.~Chen, F.~Li, C.~Li, K.~Yang, Y.~Liu, and F.~Chang, ``Channel
  measurement and modeling of the small-scale fading characteristics for urban
  inland river environment,'' \emph{{IEEE} Trans. on Wireless Commun.},
  vol.~19, no.~5, pp. 3376--3389, may 2020.

\bibitem{Yu2020a}
Y.~Yu, W.-J. Lu, Y.~Liu, and H.-B. Zhu, ``Neural-network-based root mean delay
  spread model for ubiquitous indoor internet-of-things scenarios,''
  \emph{{IEEE} Internet of Things Journal}, vol.~7, no.~6, pp. 5580--5589, jun
  2020.

\bibitem{Prokes2019}
A.~Prokes, T.~Mikulasek, M.~Waldecker, B.~K. Engiz, and J.~Blumenstein,
  ``Multipath propagation analysis for static urban environment at 60 {GHz},''
  in \emph{2019 International Conference on Electrical and Computing
  Technologies and Applications ({ICECTA})}.\hskip 1em plus 0.5em minus
  0.4em\relax {IEEE}, nov 2019.

\bibitem{Varela2001}
M.~Varela and M.~Sanchez, ``{RMS} delay and coherence bandwidth measurements in
  indoor radio channels in the {UHF} band,'' \emph{{IEEE} Trans on Veh.
  Technol.}, vol.~50, no.~2, pp. 515--525, mar 2001.

\bibitem{Steinboeck2013}
G.~{Steinb\"ock}, T.~{Pedersen}, B.~H. {Fleury}, W.~{Wang}, and R.~{Raulefs},
  ``Distance dependent model for the delay power spectrum of in-room radio
  channels,'' \emph{IEEE Trans. on Antennas and Propag.}, vol.~61, no.~8, pp.
  4327--4340, Aug 2013.

\bibitem{Pedersen2018}
T.~Pedersen, ``Modeling of path arrival rate for in-room radio channels with
  directive antennas,'' \emph{IEEE Trans. Antennas Propag.}, vol.~66, no.~9,
  pp. 4791--4805, Sep 2018.

\bibitem{Pedersen2019}
------, ``Stochastic multipath model for the in-room radio channel based on
  room electromagnetics,'' \emph{IEEE Trans. Antennas Propag.}, vol.~67, no.~4,
  pp. 2591--2603, Apr 2019.

\bibitem{Kotz2000}
S.~Kotz, N.~Balakrishnan, and N.~L. Johnson, \emph{Continuous Multivariate
  Distributions}.\hskip 1em plus 0.5em minus 0.4em\relax John Wiley {\&} Sons,
  Inc., apr 2000.

\bibitem{Nelsen2007}
\BIBentryALTinterwordspacing
R.~B. Nelsen, \emph{An Introduction to Copulas}.\hskip 1em plus 0.5em minus
  0.4em\relax Springer-Verlag GmbH, 2007. [Online]. Available:
  \url{https://www.ebook.de/de/product/5270140/roger_b_nelsen_an_introduction_to_copulas.html}
\BIBentrySTDinterwordspacing

\bibitem{Gustafson2016}
C.~Gustafson, D.~Bolin, and F.~Tufvesson, ``Modeling the polarimetric mm-wave
  propagation channel using censored measurements,'' in \emph{2016 Global
  Commun. Conf.}\hskip 1em plus 0.5em minus 0.4em\relax {IEEE}, Dec 2016.

\bibitem{Fryziel2002}
\BIBentryALTinterwordspacing
M.~Fryziel, C.~Loyez, L.~Clavier, N.~Rolland, and P.~A. Rolland, ``Path-loss
  model of the 60 {GHz} indoor radio channel,'' \emph{Microw. and Opt Technol.
  Letters}, vol.~34, no.~3, pp. 158--162, 2002. [Online]. Available:
  \url{https://onlinelibrary.wiley.com/doi/abs/10.1002/mop.10402}
\BIBentrySTDinterwordspacing

\bibitem{AAUIndustryMea}
M.~{Razzaghpour}, R.~{Adeogun}, I.~{Rodriguez}, G.~{Berardinelli}, R.~S.
  {Mogensen}, T.~{Pedersen}, P.~{Mogensen}, and T.~B. {Sørensen},
  ``Short-range {UWB} wireless channel measurement in industrial
  environments,'' in \emph{2019 Int. Conf. on Wireless and Mobile Comput.,
  Netw. and Commun. (WiMob)}, Oct 2019, pp. 1--6.

\bibitem{AyushEUCAP}
A.~Bharti, L.~Clavier, and T.~Pedersen, ``Joint statistical modeling of
  received power, mean delay, and delay spread for indoor wideband radio
  channels,'' in \emph{Eur. Conf. on Antennas and Propag.}, 2020.

\bibitem{8901446}
A.~W. {Mbugua}, W.~{Fan}, K.~{Olesen}, X.~{Cai}, and G.~F. {Pedersen},
  ``Phase-compensated optical fiber-based ultrawideband channel sounder,''
  \emph{IEEE Trans. on Microw. Theory and Techn.}, vol.~68, no.~2, pp.
  636--647, Feb 2020.

\bibitem{8713575}
X.~{Cai} and W.~{Fan}, ``A complexity-efficient high resolution propagation
  parameter estimation algorithm for ultra-wideband large-scale uniform
  circular array,'' \emph{IEEE Trans. on Commun.}, vol.~67, no.~8, pp.
  5862--5874, Aug 2019.

\bibitem{Akaike1974}
H.~Akaike, ``A new look at the statistical model identification,'' \emph{{IEEE}
  Trans. on Autom. Control}, vol.~19, no.~6, pp. 716--723, Dec 1974.

\bibitem{Claeskens2008}
G.~Claeskens and N.~L. Hjort, \emph{{Model Selection and Model
  Averaging}}.\hskip 1em plus 0.5em minus 0.4em\relax Cambridge University
  Press, 2008.

\bibitem{Ding2018}
J.~Ding, V.~Tarokh, and Y.~Yang, ``Model selection techniques: An overview,''
  \emph{{IEEE} Signal Process. Mag.}, vol.~35, no.~6, pp. 16--34, Nov 2018.

\bibitem{Rice1994}
J.~A. Rice, \emph{{Mathematical Statistics and Data Analysis}}, 2nd~ed.\hskip
  1em plus 0.5em minus 0.4em\relax Duxbury Press, 1994.

\bibitem{Efron1994}
B.~Efron and R.~Tibshirani, \emph{An Introduction to the Bootstrap}.\hskip 1em
  plus 0.5em minus 0.4em\relax Chapman and Hall/{CRC}, May 1994.

\end{thebibliography}
\bibliographystyle{IEEEtran}

\end{document}